\documentclass[conference]{IEEEtran}
\makeatletter
\def\ps@headings{%
\def\@oddhead{\mbox{}\scriptsize\rightmark \hfil \thepage}%
\def\@evenhead{\scriptsize\thepage \hfil \leftmark\mbox{}}%
\def\@oddfoot{}%
\def\@evenfoot{}}
\makeatother
\usepackage{lipsum, etoolbox, xcolor, graphicx}
\usepackage[acronym]{glossaries}
\usepackage{geometry}

\newacronym{BSec}{BeamSec}{BeamSec}
\usepackage{subcaption}
\usepackage{algorithm}
\usepackage{algpseudocode}
\usepackage{gensymb}
\usepackage{amsmath,amssymb,amsfonts,bm}

\usepackage{gensymb}
\usepackage{verbatim} 
\usepackage{multirow}
\usepackage{tabularx}
\usepackage{booktabs}
\usepackage{float}

\DeclareUnicodeCharacter{223C}{~}

\usepackage{stfloats}
\usepackage{url}
\usepackage{soul}
\usepackage[normalem]{ulem}
\soulregister\cite7 
\soulregister\citep7 
\soulregister\citet7 
\soulregister\ref7 
\soulregister\pageref7 

\usepackage{graphicx}
\usepackage{textcomp}
\usepackage{xcolor}
\usepackage{verbatim}
\usepackage{xcolor}
\usepackage{array}
\usepackage{rotating, graphicx}
\usepackage{wrapfig}
\usepackage{multirow}
\usepackage{tabularx}
\usepackage{booktabs}
\usepackage{algorithmicx}
\usepackage{algpseudocode}
\algnewcommand\algorithmicswitch{\textbf{switch}}
\algnewcommand\algorithmiccase{\textbf{case}}
\algnewcommand\algorithmicassert{\texttt{assert}}
\algnewcommand\Assert[1]{\State \algorithmicassert(#1)}%
\algdef{SE}[SWITCH]{Switch}{EndSwitch}[1]{\algorithmicswitch\ #1\ \algorithmicdo}{\algorithmicend\ \algorithmicswitch}%
\algdef{SE}[CASE]{Case}{EndCase}[1]{\algorithmiccase\ #1}{\algorithmicend\ \algorithmiccase}%
\algtext*{EndSwitch}%
\algtext*{EndCase}%

\newcommand{\bx}{\mathbf{x}}
\newcommand{\by}{\mathbf{y}}
\newcommand{\bH}{\mathbf{H}}
\newcommand{\bff}{\mathbf{f}}
\newcommand{\bw}{\mathbf{w}}
\newcommand{\bEta}{\boldsymbol{\eta}}
\newcommand{\bI}{\mathbf{I}}
\newcommand{\bzero}{\boldsymbol{0}}
\newcommand{\ba}{\mathbf{a}}
\newcommand{\bp}{\mathbf{p}}
\newcommand{\bg}{\mathbf{g}}

\newcommand{\Cset}{\mathbb{C}}
\newcommand{\Herm}{\mathsf{H}}
\newcommand{\Trans}{\mathsf{T}}
\newcommand{\Ex}{\mathsf{E}}

\newcommand{\Pset}{\mathcal{P}}
\newcommand{\Hset}{\mathcal{H}}
\newcommand{\Wcal}{\mathcal{W}}
\newcommand{\Fcal}{\mathcal{F}}
\newcommand{\Gcal}{\mathcal{G}}
\newcommand{\figref}[1]{\figurename~\ref{#1}}

\definecolor{DarkGreen}{RGB}{0,150,0}

\usepackage[font=footnotesize]{caption} 


\newcommand{ \red} [ 1]{{\color{red}#1}}
\newacronym{tx}{TX}{transmitter}
\newacronym{rx}{RX}{receiver}
\newacronym{ap}{AP}{access point}
\newacronym{adc}{ADC}{Analog to Digital Converter}
\newacronym{acp}{ACP}{Angular channel profile}
\newacronym{5g}{5G}{5th Generation}
\newacronym{aimd}{AIMD}{Additive Increase Multiplicative Decrease}
\newacronym{am}{AM}{Acknowledged Mode}
\newacronym{att1}{Quasi-omni attacker}{Quasi-omni Attacker}
\newacronym{att2}{Directional attacker}{Directional Attacker}
\newacronym{att3}{Colluding attackers}{Colluding attackers}
\newacronym{amc}{AMC}{Adaptive Modulation and Coding}
\newacronym{aqm}{AQM}{Active Queue Management}
\newacronym{awgn}{AGWN}{Additive White Gaussian Noise}
\newacronym{balia}{BALIA}{Balanced Link Adaptation}
\newacronym{bdp}{BDP}{Bandwidth-Delay Product}
\newacronym{bf}{BF}{Beamforming}
\newacronym{cc}{CC}{Congestion Control}
\newacronym{cdf}{CDF}{Cumulative Distribution Function}
\newacronym{ci}{CI}{Close-in free space reference}
\newacronym{cn}{CN}{Core Network}
\newacronym{cqi}{CQI}{Channel Quality Information}
\newacronym{cp}{CP}{Control Plane}
\newacronym{csirs}{CSI-RS}{Channel State Information - Reference Signal}
\newacronym{csi}{CSI}{Channel State Information}
\newacronym{dc}{DC}{Dual Connectivity}
\newacronym{dce}{DCE}{Direct Code Execution}
\newacronym{dci}{DCI}{Downlink Control Information}
\newacronym{dl}{DL}{Downlink}
\newacronym{dmr}{DMR}{Deadline Miss Ratio}
\newacronym{dmrs}{DMRS}{DeModulation Reference Signal}
\newacronym{e2e}{E2E}{End-to-End}
\newacronym{ecn}{ECN}{Explicit Congestion Notification}
\newacronym{edf}{EDF}{Earliest Deadline First}
\newacronym{enb}{eNB}{evolved Node Base}
\newacronym{epc}{EPC}{Evolved Packet Core}
\newacronym{es}{ES}{Edge Server}
\newacronym{fdma}{FDMA}{Frequency Division Multiple Access}
\newacronym{fdd}{FDD}{Frequency Division Duplexing}
\newacronym[firstplural=Radio Access Technologies (RATs)]{rat}{RAT}{Radio Access Technology}
\newacronym{fs}{FS}{Fast Switching}
\newacronym{ftp}{FTP}{File Transfer Protocol}
\newacronym{bs}{BS}{Base Station}
\newacronym{gnb}{gNB}{Next Generation Node Base}
\newacronym{harq}{HARQ}{Hybrid Automatic Repeat reQuest}
\newacronym{hetnet}{HetNet}{Heterogeneous Network}
\newacronym{hh}{HH}{Hard Handover}
\newacronym{hol}{HOL}{Head-of-Line}
\newacronym{ia}{IA}{Initial Access}
\newacronym{imt}{IMT}{International Mobile Telecommunication}
\newacronym{iot}{IoT}{Internet of Things}
\newacronym{los}{LoS}{Line of Sight}
\newacronym{lte}{LTE}{Long Term Evolution}
\newacronym{m2m}{M2M}{Machine to Machine}
\newacronym{mac}{MAC}{Medium Access Control}
\newacronym{mc}{MC}{Multi-Connectivity}
\newacronym{mcs}{MCS}{Modulation and Coding Scheme}
\newacronym{mec}{MEC}{Mobile Edge Cloud}
\newacronym{mi}{MI}{Mutual Information}
\newacronym{mimo}{MIMO}{Multiple Input, Multiple Output}
\newacronym{mmwave}{mmWave}{millimeter wave}
\newacronym{mr}{MR}{Maximum Rate}
\newacronym{mss}{MSS}{Maximum Segment Size}
\newacronym{mtd}{MTD}{Machine-Type Device}
\newacronym{mtu}{MTU}{Maximum Transmission Unit}
\newacronym{nsf}{NSF}{National Science Foundation}
\newacronym{nfv}{NFV}{Network Function Virtualization}
\newacronym{nlos}{nLoS}{Non-line of Sight}
\newacronym{nr}{NR}{New Radio}
\newacronym{ofdm}{OFDM}{Orthogonal Frequency Division Multiplexing}
\newacronym{pdcch}{PDCCH}{Physical Downlonk Control Channel}
\newacronym{pdcp}{PDCP}{Packet Data Convergence Protocol}
\newacronym{pdsch}{PDSCH}{Physical Downlink Shared Channel}
\newacronym{pdp}{PDP}{power delay profile}
\newacronym{pdu}{PDU}{Packet Data Unit}
\newacronym{pf}{PF}{Proportional Fair}
\newacronym{pgw}{PGW}{Packet Gateway}
\newacronym{phy}{PHY}{Physical}
\newacronym{pbch}{PBCH}{Physical Broadcast Channel}
\newacronym[plural=\gls{mme}s,firstplural=Mobility Management Entities (MMEs)]{mme}{MME}{Mobility Management Entity}
\newacronym{prb}{PRB}{Physical Resource Block}
\newacronym{pss}{PSS}{Primary Synchronization Signal}
\newacronym{pucch}{PUCCH}{Physical Uplink Control Channel}
\newacronym{pusch}{PUSCH}{Physical Uplink Shared Channel}
\newacronym{rach}{RACH}{Random Access Channel}
\newacronym{ran}{RAN}{Radio Access Network}
\newacronym{red}{RED}{Random Early Detection}
\newacronym{rf}{RF}{Radio Frequency}
\newacronym{rlc}{RLC}{Radio Link Control}
\newacronym{rlf}{RLF}{Radio Link Failure}
\newacronym{rrc}{RRC}{Radio Resource Control}
\newacronym{rrm}{RRM}{Radio Resource Management}
\newacronym{rr}{RR}{Round Robin}
\newacronym{rs}{RS}{Remote Server}
\newacronym{rsrp}{RSRP}{Reference Signal Received Power}
\newacronym{rss}{RSS}{Received Signal Strength}
\newacronym{rtt}{RTT}{Round Trip Time}
\newacronym{rw}{RW}{Receive Window}

\newacronym{sa}{SA}{standalone}
\newacronym{sack}{SACK}{Selective Acknowledgment}
\newacronym{sap}{SAP}{Service Access Point}
\newacronym{sch}{SCH}{Secondary Cell Handover}
\newacronym{scoot}{SCOOT}{Split Cycle Offset Optimization Technique}
\newacronym{sls}{SLS}{sector level sweep}
\newacronym{brp}{BRP}{beam-refinement phase} 
\newacronym{sdma}{SDMA}{Spatial Division Multiple Access}
\newacronym{sdr}{SDR}{Software-Defined Radio}
\newacronym{sinr}{SINR}{Signal to Interference plus Noise Ratio}
\newacronym{sm}{SM}{Saturation Mode}
\newacronym{snr}{SNR}{Signal to Noise Ratio}
\newacronym{son}{SON}{Self-Organizing Network}
\newacronym{ss}{SS}{Synchronization Signal}
\newacronym{srs}{SRS}{Sounding Reference Signal}
\newacronym{sss}{SSS}{Secondary Synchronization Signal}
\newacronym{sta}{STA}{station}
\newacronym{tb}{TB}{Transport Block}
\newacronym{tcp}{TCP}{Transmission Control Protocol}
\newacronym{tdoa}{TDoA}{Time difference of arrival}
\newacronym{tdd}{TDD}{Time Division Duplexing}
\newacronym{tdma}{TDMA}{Time Division Multiple Access}
\newacronym{tfl}{TfL}{Transport for London}
\newacronym{thz}{THz}{Terahertz}
\newacronym{tm}{TM}{Transparent Mode}
\newacronym{trp}{TRP}{Transmitter Receiver Pair}
\newacronym{tti}{TTI}{Transmission Time Interval}
\newacronym{ttt}{TTT}{Time-to-Trigger}
\newacronym{ue}{UE}{User Equipment}
\newacronym{ul}{UL}{Uplink}
\newacronym{uml}{UML}{Unified Modeling Language}
\newacronym{um}{UM}{Unacknowledged Mode}
\newacronym{utc}{UTC}{Urban Traffic Control}
\newacronym{vm}{VM}{Virtual Machine}
\newacronym{rsrq}{RSRQ}{Reference Signal Received Quality}
\newacronym{rssi}{RSSI}{Received Signal Strength Indicator}
\newacronym{crs}{CRS}{Cell Reference Signal}
\newacronym{comp}{CoMP}{Coordinated Multi-Point}
\newacronym{cran}{C-RAN}{Cloud \acrlong{ran}}
\newacronym{ca}{CA}{Carrier Aggregation}
\newacronym{cco}{CC}{Carrier Component}
\newacronym{nsa}{NSA}{Non Stand Alone}
\newacronym{embb}{eMBB}{Enhanced Mobility Broadband}
\newacronym{bsr}{BSR}{Buffer Status Report}
\newacronym{srb}{SRB}{Service Radio Bearer}
\newacronym{scm}{SCM}{Spatial Channel Model}
\newacronym{sctp}{SCTP}{Stream Control Transmission Protocol}
\newacronym{mptcp}{MPTCP}{Multi-path TCP}
\newacronym{ietf}{IETF}{Internet Engineering Task Force}
\newacronym{os}{OS}{Operating System}
\newacronym{tls}{TLS}{Transport Layer Security}
\newacronym{rfc}{RFC}{Request for Comments}
\newacronym{http}{HTTP}{HyperText Transfer Protocol}
\newacronym{nat}{NAT}{Network Address Translation}
\newacronym{api}{API}{Application Programming Interface}
\newacronym{rto}{RTO}{Retransmission Timeout}
\newacronym{psc}{PSC}{Public Safety Communication}
\newacronym{rpgm}{RPGM}{Reference Point Group Mobility}
\newacronym{ic}{IC}{Incident Command}
\newacronym{rsu}{RSU}{Road Side Unit}
\newacronym{uav}{UAV}{Unmanned Aerial Vehicle}
\newacronym{usa}{U.S.}{United States}
\newacronym{vr}{VR}{Virtual Reality}
\newacronym{iab}{IAB}{Integrated Access and Backhaul}
\newacronym{wlan}{WLAN}{Wireless Local Area Network}
\newacronym{cots}{COTS}{Commercial Off-the-Shelf}
\newacronym{fpga}{FPGA}{Field Programmable Gate Array}
\newacronym{rcn}{RCN}{Research Coordination Network}
\newacronym{abg}{ABG}{Alpha-Beta-Gamma}
\newacronym{fi}{FI}{Floating Intercept}
\newacronym{uas}{UAS}{Unmanned Aerial System}
\newacronym{gps}{GPS}{Global Positioning System}
\newacronym{a2g}{A2G}{air-to-ground}
\newacronym{a2a}{A2A}{air-to-air}
\newacronym{uma}{UMa}{Urban Macro}
\newacronym{umi}{UMi}{Urban Micro}
\newacronym{rma}{RMa}{Rural Macro}
\newacronym{inoo}{InOo}{Indoor Open Office}
\newacronym{ple}{PLE}{path loss exponent}
\newacronym{aoa}{AoA}{Angle of Arrival}
\newacronym{aod}{AoD}{Angle of Departure}
\newacronym{toa}{ToA}{Time of Arrival}
\newacronym{mpc}{MPC}{Multi-path Component}
\newacronym{cir}{CIR}{Channel impulse response}
\newacronym{rt}{RT}{Ray-tracing}
\newacronym{tc}{TC}{Time Cluster}
\newacronym{sl}{SL}{Spatial Lobe}
\newacronym{6g}{6G}{Sixth Generation}
\newacronym{ns3}{ns-3}{Network Simulator 3}
\newacronym{fsc}{FS}{Fully Stochastic}
\newacronym{hbc}{HB}{Hybrid}
\newacronym{hpbw}{HPBW}{Half Power Beamwidth}
\newacronym{hsc}{HSC}{Hybrid Semantic Compression}
\newacronym{prr}{PRR}{Packet Receipt Rate}

\def\BibTeX{{\rm B\kern-.05em{\sc i\kern-.025em b}\kern-.08em
    T\kern-.1667em\lower.7ex\hbox{E}\kern-.125emX}}
    \usepackage[affil-it]{authblk}
\begin{document}

\date{}

\title{BeamSec: A Practical mmWave Physical Layer Security Scheme Against Strong Adversaries}

\author{Afifa Ishtiaq}
\author{Arash Asadi}
\author{Ladan Khaloopour}
\author{Waqar Ahmed}
\author{Vahid Jamali}
\author{Matthias Hollick}
\affil{\textit{Technische Universität Darmstadt}}
\affil{\textit{\{afifa.ishtiaq,arash.asadi,ladan.khaloopour,waqar.ahmed1,vahid.jamali,matthias.hollick\}@tu-darmstadt.de}}

\maketitle

\begin{abstract}
The high directionality of millimeter-wave (mmWave) communication systems has proven effective in reducing the attack surface against eavesdropping, thus improving the physical layer security. However, even with highly directional beams, the system is still exposed to eavesdropping against adversaries located within the main lobe. In this paper, we propose \acrshort{BSec}, a solution to protect the users even from adversaries located in the main lobe. The key feature of \acrshort{BSec} are: $(i)$ Operating without the knowledge of eavesdropper's location/channel; $(ii)$ Robustness against colluding eavesdropping attack and $(iii)$ Standard compatibility, which we prove using experiments via our IEEE 802.11ad/ay-compatible 60 GHz phased-array testbed. Methodologically,  \acrshort{BSec} first identifies uncorrelated and diverse beam-pairs between the transmitter and receiver by analyzing signal characteristics available through standard-compliant procedures. Next, it encodes the information jointly over all selected beam-pairs to minimize information leakage. We study two methods for allocating transmission time among different beams, namely uniform allocation (no knowledge of the wireless channel) and optimal allocation for maximization of the secrecy rate (with partial knowledge of the wireless channel). Our experiments show that \acrshort{BSec} outperforms the benchmark schemes against single and colluding eavesdroppers and enhances the secrecy rate by 79.8\% over a random paths selection benchmark. 

\end{abstract}



\section{Introduction}





Practical physical layer security, utilizing wireless channel randomness, has been a vibrant research area for over a decade, offering a complementary approach to higher-layer cryptography. The physical layer security research is divided into two categories: $(i)$ key generation: deriving a secure key from the channel parameters of the communicating parties to eliminate the need for high-layer key exchange and authenticating the users via these parameters \cite{24,25}, and $(ii)$ protection against eavesdropping: to reduce the \gls{sinr} at the eavesdropper's location such that the packets can no longer be decoded \cite{2,32}. This is based on reducing the {\it signal footprint}, i.e., the physical area where the transmitted signal is heard. Common approaches rely on the transmission of artificial noise in the direction of eavesdropper~\cite{31}, tighter beamforming towards the receiver~\cite{26}, and side-lobe reduction~\cite{32}.  

\subsection{Motivation}
\begin{figure}
\centering
    \subfloat[]{
    \includegraphics[width=.45\linewidth]{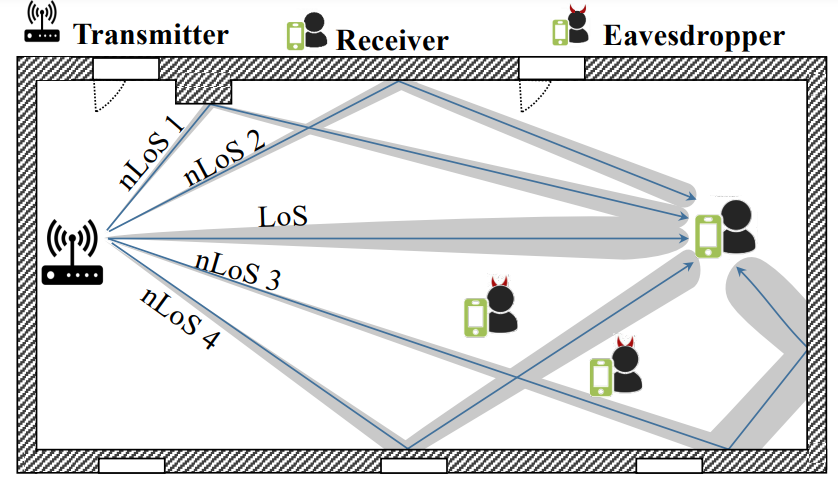}
    \label{fig_overviewa}
    }
    \subfloat[]{
    \includegraphics[width=.42\linewidth]{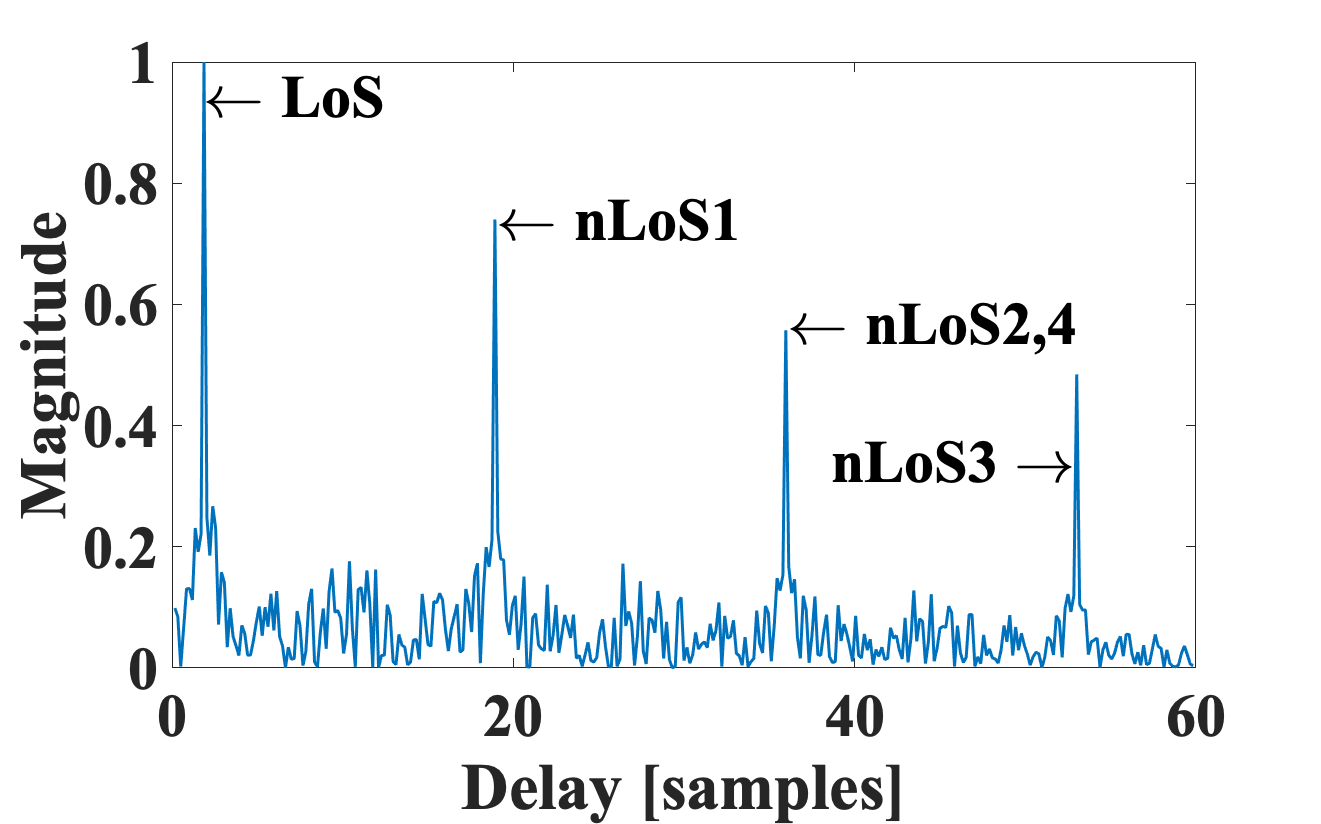}
    \label{fig_overviewb}
    }
   \caption{Path selection challenges: (a) The paths between TX and RX in an indoor mmWave scenario; (b) Power delay profile (simulated) for (a).}
\end{figure}

The advent of millimeter-wave (mmWave) systems with phased-array antennas has allowed the use of highly directional beams, thus minimizing the signal footprint. However, even at such systems, the adversaries can still eavesdrop on the non-negligible side-lobes of consumer-grade antennas, which have been the subject of several studies~\cite{28}, \cite{35}. {\it However, the main lobe remains exposed to adversaries, see \figref{fig_overviewa}}. Prior work often argues that an attacker on the main-lobe can be easily detected. In practice, the small size of consumer-grade wireless devices allows eavesdropping without creating (easily) detectable RF signature. 
Further, knowledge of eavesdroppers' location/channel is central to the majority of existing work~\cite{9,10,8}. This assumption undermines the practicality of such approaches since obtaining such information is very difficult in a real system with non-cooperative adversaries. We believe there exists a gap in solutions addressing main-lobe security under practical constraints (e.g., knowledge of eavesdroppers and hardware capabilities). 
 This calls for a new design exploiting the standard mmWave communication procedures to provide practical physical layer security at mmWave networks.

\subsection{Related work}
There exist only a few works addressing main-lobe security. The works in~\cite{9,10,8} secure the main lobe through rotated angular beamforming with a frequency-diverse array assuming the knowledge of the eavesdropper's location. In ~\cite{11}, the authors exploit antenna arrays and ground reflections to achieve angle-range-dependent transmission and side-lobe randomization for location-based physical layer security in mmWave vehicular networks. In \cite{12}, the authors propose a dual-beam transmission technique that ensures the main-lobe is coherent only at the legitimate \gls{rx}’s location. The work in \cite{13} proposes an artificial-noise-aided  hybrid precoder that maximizes the secrecy rate assuming full channel knowledge of the eavesdropper. The work in ~\cite{15a} proposes an optimal directional modulation with artificial noise  using a frequency-diverse phased array scheme to decouple the angle-range correlation and maximize the secrecy rate of \gls{rx}. 

Although a big step forward, the above-mentioned solutions: $(i)$ rely on specialized antennas (angular polarization)~\cite{12}, or a large number of RF chains~\cite{15a}; $(ii)$ require the exact knowledge of eavesdropper's location~\cite{9,10,8} or channel~\cite{13}; or $(iii)$ introduce additional interference to the network by transmitting artificial noise~\cite{14a}. These limitations pose a major obstacle in the practical applications of physical layer security in  mmWaves systems. In \cite{29,16,30}, the authors propose to perform path/beam hopping between randomly chosen paths between transmitter and receiver.  Although theoretically effective, {\it beam hopping without rigorous analysis of the correlation among \gls{tx}-\gls{rx} beams can adversely impact the security}: $(i)$ by increasing the signal footprint, $(ii)$ a random selection of paths may lead to selecting the paths with minimum angular separation or minimum diversity, which diminishes the essence of hopping. To the best of the authors' knowledge, the \textit{optimized selection} of the diverse but potentially  uncorrelated 
channel paths for the maximization of the secrecy rate has not been investigated in the literature yet. 

\subsection{Our proposal: \acrshort{BSec}}

In this article, we propose a practical scheme, named \acrshort{BSec}, to reduce the attack surface area and information leakage by an optimal splitting of the data streams across different paths/beams between the legitimate transceivers. 

{\bf Challenges.} 
As mentioned, splitting user data over multiple paths toward the receiver can theoretically reduce the probability of an eavesdropper intercepting the entire data. However, simple transmission over any path does not necessarily enhance security. {\it Thus, finding a practical method to identify the components resulting in higher security without exact knowledge of the eavesdropper's location is a challenge.} 
Measurement campaigns have revealed  mmWave channels are sparse (the number of paths between \gls{tx} and \gls{rx} are limited)~\cite{rappaport201238}  and there is a spatial correlation correlated among these paths even in the presence of blockages~\cite{sur2016beamspy}. This makes {\it secure beam hopping} at mmWave challenging. \textit{To enhance the physical layer security, it is crucial to identify the paths that are spatially diverse and distinct.} From a channel perspective, two paths are considered diverse and distinct if they have different delays and diverse \gls{aod}. Therefore, spatially distinct and diverse paths with non-overlapping coverage areas are essential for enhancing  physical layer security. For example, in \figref{fig_overviewa}, \gls{nlos} 1 and \gls{nlos} 2 have similar coverage areas, while \gls{nlos} 2 and \gls{nlos} 4 are non-overlapping but indistinguishable in delay profile. Such ambiguities can be resolved using high-end channel sounders but pose challenges for practical solutions within commercial hardware limitations.

 \begin{figure}[t]
\centering
\includegraphics[scale=0.355]{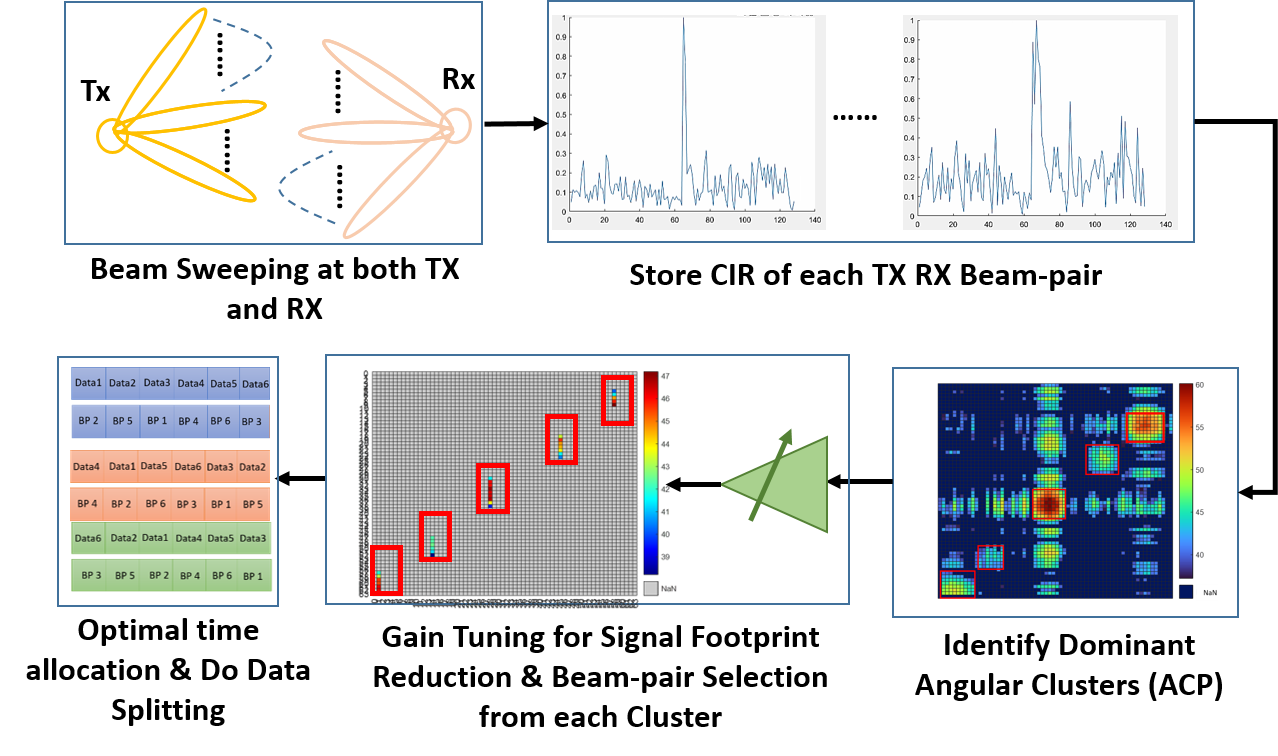}
\caption{An overview of the \acrshort{BSec} procedure.}
\label{fig: Beamselection}
\end{figure}

{\bf Overview and contributions.}
\acrshort{BSec} is a low complexity real-time physical layer security technique that enhances main-lobe security. \acrshort{BSec} is effective against passive eavesdroppers whose location and channel information are completely unknown. This is achieved through a practical method for distinguishing distinct and diverse paths by analyzing signal characteristics (i.e., \gls{aoa}, \gls{aod}, and \gls{rssi}) which can be obtained by standard 802.11ad procedures. After identifying the paths, we propose a fast linear optimization method to maximize the overall secrecy rate based on the coverage area of each path. \figref{fig: Beamselection} shows the overview of \acrshort{BSec}. Below is the summary of our main contributions:
\begin{itemize}

\item To the best of our knowledge, this is the first work on main-lobe physical layer security of mmWave system that does not require the knowledge of eavesdroppers' location/channel, use of specialized antennas~\cite{12}, or digital/hybrid beamforming. 

\item We propose \acrshort{BSec}, a method for identifying diverse and distinct beam-pairs using signal characteristics extracted from the 802.11ad beam-training procedure. Furthermore, we reduce the signal footprint by adapting RF and beamforming  gain for each beam-pair. Without the knowledge of the wireless channel, \acrshort{BSec} splits the data `equally' among the selected beam-pairs.

\item Next, we analytically model the resulting information leakage and absolute secrecy rate. Leveraging our model, we develop a method that can further optimize \acrshort{BSec}'s data splitting when a small subset of channel measurements in the environment is available. With this `partial channel knowledge', we optimize the time allocation to each beam-pair such that the overall secrecy rate is maximized. {\it Note that in both scenarios, the eavesdropper's location is unknown.}  



\item We experimentally evaluate \acrshort{BSec} using our 802.11ad/ay-compatible 60 GHz phased-array testbed against three strong adversary models: quasi-omni, directional, and colluding attackers. 
Experiments reveal that \acrshort{BSec} increases the secrecy rate compared to random path selection \cite{16} by 79.8\%  and $\sim\!\!31.2\%$ with no knowledge and partial channel knowledge, respectively. Moreover, \acrshort{BSec} can protect against up to six colluding eavesdroppers.



\end{itemize}

\section{Overview of IEEE 802.11{ad}}
\begin{figure}[H]
\begin{minipage}{.45\linewidth}
    \includegraphics[width=.9\linewidth,keepaspectratio]{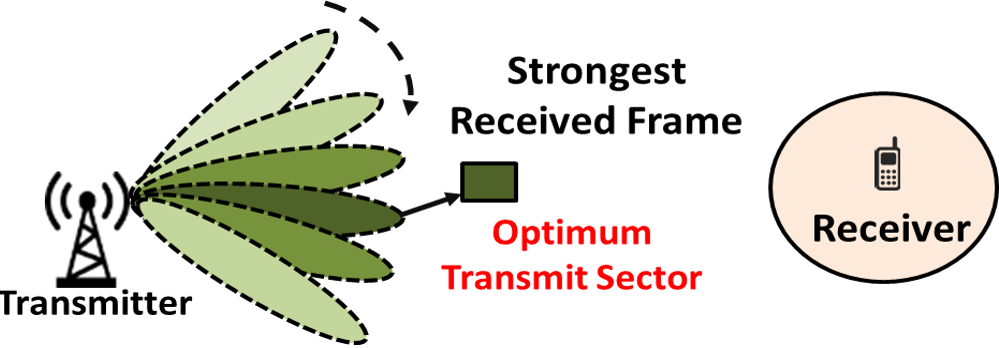}
    \subcaption{Transmit sector sweep }
    \label{Txss}
    \end{minipage}
\hfill
\begin{minipage}{.45\linewidth}

    \includegraphics[width=1\linewidth,keepaspectratio]{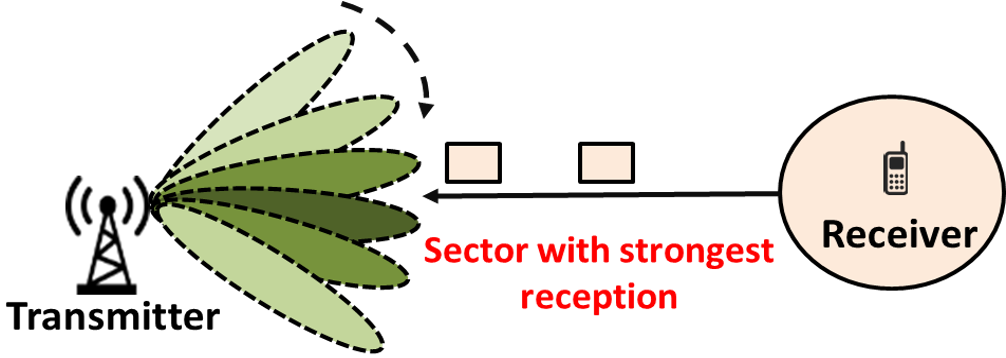}
    \subcaption{Receive sector sweep }
    \label{Rxss}
\end{minipage} 
\caption{Sector level sweep procedure of 802.11ad/ay.}
\label{fig: SLS}
\end{figure}

\begin{figure}[H]
\centering
    \begin{minipage}{.45\linewidth}
    \includegraphics[width=.8\linewidth,keepaspectratio]{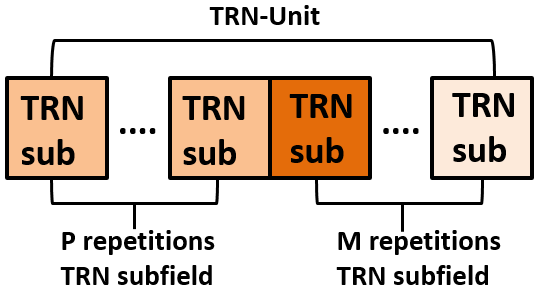}
    \subcaption{TRN representation}
    \label{BRP1}
    \end{minipage} 
     \hspace{-1em}
    \begin{minipage}{.45\linewidth}
    \includegraphics[width=1\linewidth,keepaspectratio]{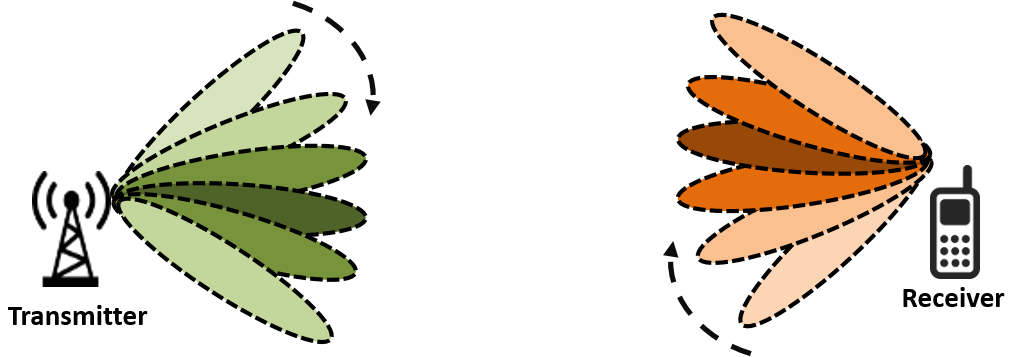}
    \subcaption{TX RX Sweep}
    \label{BRP2}
    \end{minipage} 
    \caption{Beam refinement procedure of 802.11ad/ay.}
    \label{fig: BRP}
\end{figure}

In the following, we briefly review IEEE 802.11{ad}'s beam-training, forming \acrshort{BSec}'s foundation. Commercial 802.11ad devices utilize a predefined codebook for wide and narrow beam coverage, achieving high data rates. This beam-training consists of two stages, \gls{sls} and \gls{brp}, for \gls{ap} and \gls{sta} beam selection (Figures \ref{fig: SLS} and \ref{fig: BRP}).

During \gls{sls}, a series of frames with multiple packets are exchanged between the \gls{ap} and \gls{sta} over different antenna sectors to determine the sector with the highest signal quality. The SLS consists of two types of sector sweep: the transmit sector sweep (TXSS) and the receive sector sweep (RXSS). In the TXSS, frames containing packets are transmitted over different directions in the AP's coverage using the antenna weight vectors (AWV). The STA in listening mode decodes the header and data field, associates itself with the AP, and sends feedback to ensure that the selected transmit AWV is appropriate. During RXSS, transmission on the best-known sector from TXSS allows for finding the optimal receive sector.

After identifying the optimal sector and initial configuration of AWV, the antenna settings for the \gls{ap} and \gls{sta} are further refined using \gls{brp}. Unlike \gls{sls}, \gls{brp} does not rely on predefined sector patterns but uses directional beam scanning at both \gls{ap} and \gls{sta}. \gls{ap} is set to a specific AWV, evaluated for each AWV at \gls{sta}, and repeated for all AWVs on \gls{ap}. In the end, feedback is carried out to determine the optimal transmit and receive antenna configurations for \gls{ap} and \gls{sta}, respectively. The procedure is repeated for \gls{ap} as a receiver and \gls{sta} as a transmitter and exhausts every possible combination of transmit AWVs for a fixed receive AWV setting, leading to significant performance improvement over SLS-based training.

\section{System and Adversary Model}

Our system model consists of a transmitter (i.e., Alice) communicating with a legitimate receiver (i.e., Bob) in a multi-path environment in the presence of eavesdropper(s) (Eve). Eve is passive (i.e., it does not manipulate the communication between Alice and Bob) and its location and channel is unknown. In the following, we describe the considered channel and eavesdropping models in detail.

\subsection{Channel model}

We assume Alice, Bob, and Eve are multiple antenna nodes, respectively equipped with \(N_a\), \(N_b\), and \(N_e\) antennas. The signal model is given by
\begin{equation}
	\by_i = \bH_i \bx + \bEta_i,\quad \forall i \in\{b,e\},
\end{equation}
where subscripts $b$ and $e$ denote Bob and Eve, respectively. Here, $\bx\in\Cset^{N_a}$ denotes the transmit vector satisfying the transmit power constraint $\Ex\{\bx^H\bx\}\leq P$,  $\bH_i\in \Cset^{N_i\times N_a}$ is the channel between Alice and node $i$, $\by_i\in \Cset^{N_i}$ denotes the received signal vector at node $i$, and $\bEta_i\sim\mathcal{CN}(\bzero,\sigma_\eta^2\bI)\in \Cset^{N_i}$ denotes the zero-mean additive white Gaussian noise (AWGN) at node $i$, where $\sigma_\eta^2$ is the noise power at each antenna.

The channel matrix $\bH_i$ is generally composed of a \gls{los} component $\bH_i^{LoS}$ and \gls{nlos} components $\bH_i^{nLoS}$:
\begin{equation}\label{Eq:LoSnLoS}
	\bH_i = \bH_i^{LoS}  +  \bH_i^{nLoS}.
\end{equation}
The \gls{los} channel $\bH_i^{LoS}$ between Alice and node $i$ is: 
\begin{align}\label{Eq:LoS}
	\bH_i^{LoS} &= \alpha_{i,0} \ba_i(\phi_{i,0}){\ba_a^{\Herm}(\theta_{i,0})}, \quad \forall i \in\{b,e\},
\end{align}
where $\alpha_{i,0}\in\Cset$ is the channel coefficient of the \gls{los} link,  ${\ba_a(\theta_{i,0})}\in\Cset^{N_a}$ denotes the transmit steering vector for Alice evaluated at \gls{aod} $\theta_{i,0}$, and  $\ba_i(\phi_{i,0})\in\Cset^{N_i}$ represents the receive steering vector at node $i$ evaluated at \gls{aoa} $\phi_{i,0}$. Assuming that the \gls{nlos} channel $\bH_i^{nLoS}$ between Alice and node $i$ comprises of $L_i$ paths, we have:
\begin{align}\label{Eq:nLoS}
	\bH_i^{nLoS} &= \sum_{l=1}^{L_i}\alpha_{i,l} \ba_i(\phi_{i,l}){\ba_a^{\Herm}(\theta_{i,l})}, \quad \forall i \in\{b,e\},
\end{align}
where subscript $l$ denotes the path $l$ of the \gls{nlos} link, and $\alpha_{i,l}$,  $\phi_{i,l}$, and $\theta_{i,l}$ are the corresponding channel coefficient, \gls{aoa}, and \gls{aod}, respectively. Assuming a uniform linear array (ULA) at all nodes, the steering vectors can be written as
\begin{equation}
	\ba_i(\theta)\!\! =\!\! \big[1, e^{-j\kappa d\cos(\theta)}\!\!, \dots, e^{-j\kappa(N_i-1) d\cos(\theta)}\big]^\Trans\!\!,\, \forall i \in\{a,b,e\},
\end{equation}
where $\kappa=\frac{2\pi}{\lambda}$ is the wave number, $\lambda$ is the wavelength, and $d$ is the element spacing.

\subsection{Effective beam-space model}


We assume that Alice transmits a single data stream via linear precoding/beamforming $\bw_l\in\Cset^{N_a}$, i.e., $\bx=\bw_l x$, where $x\in\Cset$ is the data symbol that satisfies the power constraint $\Ex\{|x|^2\}\leq P$ and $\bw_l$ is unit-norm, i.e., $\|\bw_l\|^2=1$, beamforming vector for the $l$-th path. Bob adopts a linear combiner $\bff_l\in\Cset^{N_b}$, which is unit norm, i.e., $\|\bff_l\|^2=1$. In practice, Alice's beamformer $\bw_l$ and Bob's combiner $\bff_l$ are chosen from predefined codebook (i.e., beams), denoted by $\Wcal$ and $\Fcal$, respectively, which account for the aforementioned beam refinement and gain tuning procedures. For transmission at path $l$, the signal at Bob's combiner $y_b\in\Cset$ is obtained as
\begin{align}
	y_b &= \bff_l^\Herm\by_b = \bff_l^\Herm\big[\bH_b \bw_l x + \bEta_b\big]. 
\end{align}
Hence, the achievable \gls{snr} for $l^{th}$ transmission, $\gamma_{b,l}$, is
\begin{align}\label{Eq:SNRbob}
	\gamma_{b,l} = \frac{P|\bff_l^\Herm\bH_b \bw_l|^2}{\sigma_{\eta}^2}.
\end{align}
For the ideal case where Bob and Alice are able to respectively generate narrow beams towards the $l^{th}$ beam-pair, i.e., $\bw_l=\frac{1}{\sqrt{N_a}}\ba_a(\theta_{b,l})$ and $\bff_l = \frac{1}{\sqrt{N_b}}\ba_b(\phi_{b,l})$,   the SNR scales with the number of antennas at Alice and Bob (or, equivalently, the TX and RX antenna gains), i.e., $\gamma_{b,l}\propto N_a N_b$.

We assume the eavesdropper's channel matrix is unknown. The signal received at Eve after receive combining is
\begin{align}
y_e=	\bg_l^H\by_e = \bg_l^H [\bH_e \bw_l x + \bEta_e],
\end{align}
where $\bg_l\in\Cset^{N_e}$ is the combiner used by Eve. The achievable SNR depends on the eavesdropping capability of Eve, which is discussed in the following subsection in detail.

\subsection{Attacker model}

Three attacker models are considered: quasi-omni, directional, and colluding attackers. The quasi-omni attacker can overhear transmissions through the \gls{los} path or reflections but lacks directionality and the decoding ability for weak signals. For an omni-directional receive beam, Eve's receive combiner is denoted as $\bg=\frac{1}{\sqrt{N_e}}[1,\dots,1]^\Trans\triangleq \frac{1}{\sqrt{N_e}}\boldsymbol{1}$.  Therefore, Eve's SNR, denoted by $\gamma_{e,l}$, is given by 
\begin{align}\label{Eq:SNRe1}
	\gamma_{e,l} = \frac{P|\boldsymbol{1}^{\Trans}\bH_e\bw_l|^2}{N_e\sigma_{\eta}^2}.
\end{align}
The directional attacker is a strong adversary capable of aligning its \gls{rx} beam toward the best direction for overhearing Alice. 
 Let the set of receive beams  of Eve be denoted by $\Gcal=[\bg_1, \bg_2, \dots,  \bg_{|\Gcal|}]$ then 
the achievable SNR for $l$-th path at Eve $e$ is given as:
\begin{align}\label{Eq:SNRe2}
	\gamma_{e,l} = \underset{\bg_l\in\Gcal}{\max}\frac{P|\bg_l^\Herm\bH_e\bw_l|^2}{\sigma_{\eta}^2}.
\end{align}
The colluding attacker is even a strong adversary model which considers both quasi-omni and directional capabilities, with each eavesdropper scanning the channel and selecting the best-receiving beam for eavesdropping.  In particular, a set of eavesdroppers denoted by a set $\mathcal{Q}={[1,2,\dots, Q]}$ scan the channel, and for each transmission, the signal  with the maximum SNR is selected for eavesdropping.
Let $\bH_l^{(q)}$ is the channel of the $q$-th eavesdropper and  $\Gcal^{(q)}$ is the set of receive beams adopted by the $q$-th eavesdropper, i.e., $\bg^{(q)}\in\Gcal^{(q)}$. This leads to the following achievable SNR
\begin{align}\label{Eq:SNRe3}
	\gamma_{e,l} = \underset{q\in{ \mathcal{Q}}}{\max}\,\,\underset{\bg_l^{(q)}\in\Gcal^{(q)}}{\max}\frac{P|(\bg_l^{(q)})^\Herm\bH_e^{(q)}\bw_l|^2}{\sigma_{\eta}^2}.
\end{align}

\section{BeamSec}

\acrshort{BSec} minimizes the area in which communication between Alice and Bob can be overheard either from the main-lobe, side-lobe, or the reflections. Instead of communicating via the best beam-pair (i.e., the best \gls{tx} beam for Alice and the best \gls{rx} beam for Bob), \acrshort{BSec} identifies a diverse set of beam-pairs to enhance security.
As summarized in Algorithm~\ref{alg:cap}, \acrshort{BSec} achieves its goal in four steps: 1) Reducing the signal footprint via TX gain tuning and \gls{aod}/\gls{aoa} analysis of \gls{acp} to select the best beam-pair; 2) time allocation among the selected beam-pairs, uniform in absence of any knowledge of the wireless channel and optimal with partial knowledge of wireless channel; 3) rate and codebook selection; and finally 4) secure communication over the beam-pairs. The description of Steps 2 and 3 depends on the specific secure communication strategy adopted in Step 4. Therefore, in contrast to their logical implementation order 1-4, we explain Step 4 before Steps 2 and 3. 


\begin{algorithm}[htbp] 
\footnotesize
	\caption{\acrshort{BSec}}
	\begin{algorithmic}[1]
            \Statex \hspace{-0.4cm}\textbf{Step~1: Signal footprint reduction and clustering via angular channel profile analysis} 
		\State	Run 802.11ad beam training at Alice and Bob.
        \State For each beam-pair $(\bw_l,\bff_{l'})$ from \gls{tx} codebook $\bw_l\in\Wcal$ and \gls{rx} codebook $\bff_{l'}\in\Fcal$, estimate the channel impulse response (CIR) $h_{l,l'}$.
              
         \State For each TX beam $l$, select the RX beam $l'$ with the highest CIR: $\mathrm{argmax}_{l'}\,\,|h_{l,l'}|$.          
         \State Gain tuning: Reduce the transmit power while ensuring $|h_{l,l'}|\geq\tau_{ACP}$.
        \State Recompute angular channel profile.
        \State Apply K-Means for clustering potential TX-RX pairs.

        \State Add decodable and diverse TX-RX pairs in $\textbf{L}$.
  
            \Statex \hspace{-0.4cm}\textbf{Step~2: Time allocation} 
              \Switch{time allocation policy}
              \Case{uniform time allocation: Set $T_l=\frac{1}{L},\,\,\forall l$.}
             \EndCase
              \Case{optimized time allocation: Compute $T_l,\,\,\forall l$ \Comment{Eq. \eqref{Eq:TlProb}}}
              \EndCase
             \EndSwitch
            \Statex \hspace{-0.4cm}\textbf{Step~3: Rate and codebook selection} 
             \Switch{rate selection policy}
              \Case{zero leakage:}
             Compute $\bar{C}_{s}^{\rm abs}$ and set $R_s\leq \bar{C}_{s}^{\rm abs}$. \Comment{Eq. \eqref{Eq:Rs_abs}}
             \EndCase
              \Case{given leakage probability:}
             Set $R_s$ that yields $P_{\rm leak}$. \Comment{Eq. \eqref{Eq:Pleak}}
              \EndCase
             \EndSwitch		
  \State	For given $R_s$ and $R_{b,l}<C_{b,l},\,\,\forall l$, construct codebooks for secure message $\mathcal{C}_s$, beam-pair transmissions $\mathcal{C}_l,\,\,\forall l$, and  $\mathcal{C}=\mathcal{C}_1\times \cdots\times\mathcal{C}_{L}$. The product codebook $\mathcal{C}$ is randomly partitioned into $2^{nR_s}$ parts.
            \Statex \hspace{-0.4cm}\textbf{Step~4: Secure communication over selected beam-pairs} 
            \Statex \hspace{-0.4cm}\textit{$\%$  Encoding:} 
		 \State For each $nR_s$ information bits, select the corresponding secure message $W_s\in \mathcal{C}_s$.
 \State For a given $W_s$, randomly select a codeword  from the $W_s$th partition in $\mathcal{C}$, which comprises $L$ codewords from codebooks $\mathcal{C}_1, \ldots,\mathcal{C}_{L}$.
 \Statex \hspace{-0.4cm}\textit{$\%$  Transmission/Reception:} 
 \For{$l=1,\dots, L$}
    \State Communicate the selected codeword from $\mathcal{C}_l$  using the corresponding \gls{tx}-\gls{rx} beam-pair $(\bw_l,\bff_{l})$.
      \EndFor
  \Statex \hspace{-0.4cm}\textit{$\%$  Decoding:} 
  \State Decode all $L$ codewords from codebooks $\mathcal{C}_1, \ldots,\mathcal{C}_{L}$, respectively.
  \State From codebook $\mathcal{C}$, identify the message $W_s$ and the corresponding bits. 
	\end{algorithmic}
   \label{alg:cap}
\end{algorithm}

%

\subsection{Signal footprint reduction and clustering}
The propagation paths of the beams from different angular directions depend strongly on the spatial environment. 
Environmental factors (e.g., position, orientation, and surface characteristics of the reflectors and blockages) are of great importance for selecting secure beam-pairs. As the first step in Algorithm \ref{alg:cap}, \acrshort{BSec} exploits the environment and spatial channel conditions by using the 802.11ad beam training procedure to identify beams that lead to high \gls{snr}. Consequently, the \gls{acp} between Alice and Bob is created using the \gls{cir} values for each beam combination obtained from the 802.11ad beam training procedure. Next, the RX beam with the best channel towards TX is selected (Line 3). This is an important step to improve the secrecy rate with minimal impact on the system's capacity.

\begin{figure}[t]
 \centering 
    \subfloat[Before gain-tuning]{
        \includegraphics[scale=0.49,keepaspectratio]{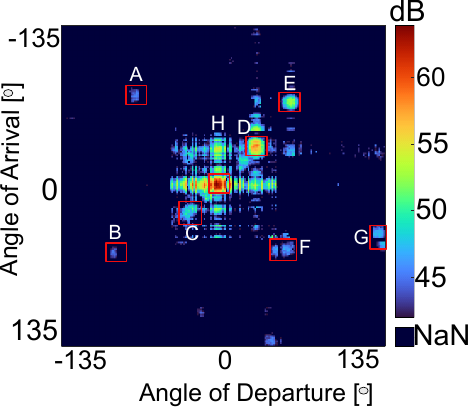}
    \label{fig: Heatmap}
    }
\subfloat[After gain-tuning]{
            \includegraphics[width=0.46\linewidth]{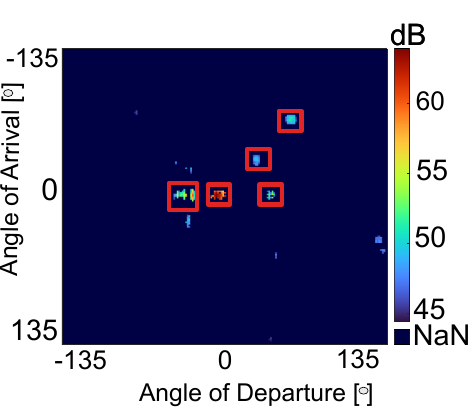}
            \label{fig:pre_gain_tuning}
        }
        
    
        \caption{Angular channel profile of the seminar room.}
        \label{fig: Gain Tuning}
\end{figure}

In \figref{fig: Heatmap}\footnote{The figure shows labeled clusters (\textbf{A-H}), with a central high-power cluster (\textbf{H}) from the \gls{los} link, and smaller clusters away due to indoor \gls{nlos} links.},
we observe that the default codebook has a large signal footprint, which results in lower secrecy rate. In \acrshort{BSec}, we exploit a gain tuning mechanism that reduces the information leakage from both main- and side-lobes. Specifically, the antenna gains are tuned such that $|h_{l,l'}|\geq\tau_{ACP}$. This is better shown in \figref{fig:pre_gain_tuning}, where we can clearly see a reduction in signal footprint in ACP after gain tuning. Next, we use K-means algorithms to perform clustering among the remaining TX-RX beam pairs. After gain tuning, we select TX-RX beam pairs that have an uncorrelated channel, i.e., each TX beam belongs to a different angular cluster and is spatially diverse from each other.

\subsection{Secure coding over refined beam-pairs}\label{Sec:Secure}


Based on the previously mentioned steps, \acrshort{BSec} develops a refined codebook with a set of beam-pairs and their respective gains, which is agreed upon between Alice and Bob. We index these beam-pairs by $l=1,\dots,L$, where $L$ is the total number of refined beam-pairs by BeamSec. The selected beam-pairs provide high channel capacity between Alice and Bob and have minimal information leakage to potential eavesdroppers.
Nonetheless, we consider a general setting where the fraction of time allocated to each beam-pair can be different and subject to optimization (see Section~\ref{Sec:TimeAllocation} for the proposed time allocation scheme). Let $T_l$ denote the fraction of time assigned to communication over path $l$, where $\sum_{l=1}^{L}T_l=1$ and $T_l\geq 0,\,\,\forall l$, have to hold. Moreover, the transmitted data is split among the total number of beam-pairs, i.e., each beam-pair is assigned to transmit and receive its specific data part (Step 4).
The secrecy is ensured by encoding information, not into the individual data segments sent over different beams, but rather by \textit{jointly} encoding information across all $L$ beam-pairs. The code construction, encoding, and decoding adopted for BeamSec are concisely summarized in lines 14-21 of Algorithm \ref{alg:cap}, which follow similar techniques as those for the general physical-layer security schemes with partial  channel knowledge, see, e.g., \cite{30} for details.

In the following, we first investigate the absolute secrecy rate for which  the information leakage to the eavesdropper(s) is zero. Subsequently, we generalize this performance measure and characterize the information leakage probability for any given fixed rate above the absolute secrecy rate. 

\subsubsection{Absolute secrecy rate}
Eve's \gls{snr} $\gamma_{e,l}$ depends on the channel matrix $\bH_e$, which is  unknown and depends on its location. Let $\bp_e\in\Pset_e$ denote Eve's location where $\Pset_e$ denotes all possible Eve locations.  A given hypothesis on $\bp_e$ gives us certain knowledge about $\bH_e$. For example, if $\bp_e$ is close to Alice's location or is along the direction that the beamformer $\bw_l$ targets, then we expect a stronger Eve's channel $\bH_e$ and hence a larger $\gamma_{e,l}$. To formalize this concept, we assume that $\bH_{e}\in\Hset_{e}(\bp_e)$, where $\Hset_{e}(\bp_e)$ denotes the set of all possible Eve's channel matrices $\bH_{e}$ with structure in \eqref{Eq:LoSnLoS}-\eqref{Eq:nLoS} given that Eve's location is $\bp_e$.  Based on this notation,  the  achievable absolute secrecy rate (i.e., zero information leakage) across all $L$ transmissions can be obtained as
\begin{align}\label{Eq:Rs_abs}
	\bar{C}_{s}^{\rm abs} = \underset{\bp_e\in\Pset_e}{\min} &\sum_{l=1}^{L} T_l C_{s,l}(\bp_e),
\end{align}
with  
$
C_{s,l}(\bp_e) = \big[C_{b,l} - \underset{\bH_e\in\Hset_{e}(\bp_e)}{\max} C_{e,l}(\bH_e)\big]^+
$,
where $C_{b,l}=\log_2(1\!+\gamma_{b,l})$, $C_{e,l}(\bH_e)=\log_2(1+\gamma_{e,l}(\bH_e))$, and $[z]^+\triangleq \max(0,z)$.
The operator $\underset{\bH_e\in\Hset_{e}(\bp_e)}{\max}$ corresponds to the worst-case channel realization given that Eve is at position $\bp_e$ and the operator $\underset{\bp_e\in\Pset_e}{\min}$ accounts for the worst-case location of Eve (in terms of secrecy rate). Thus, the secrecy rate in \eqref{Eq:Rs_abs} accounts for all possible Eve's locations and {\it does not rely on the knowledge of actual Eve's location}. Hence, $R_s\leq \bar{C}_{s}^{\rm abs}$ guarantees that no information is leaked to Eve, regardless of her location and the instantaneous channel quality.  

\subsubsection{Information leakage probability}
The secrecy rate in  \eqref{Eq:Rs_abs} requires the quantification of worst-case Eve's rate for each Eve's location $\bp_e$, i.e., $\underset{\bH_e\in\Hset_{e}(\bp_e)}{\max}\,\, C_{e,l}(\bH_e)$. This quantity is difficult to analyze analytically since it needs assumptions about the wireless channel, i.e., $\Hset_{e}(\bp_e)$. In practice, it can be obtained based on empirical measurements if sufficient measurements on different locations of the environment $\bp$ have been collected by the \textit{legitimate receiver}, which can be used as an approximation for Eve's channel too. On the other hand, the value of the absolute secrecy rate in  \eqref{Eq:Rs_abs} can be quite small due to some worst-case Eve's channel conditions that occur extremely rarely. Hence, the absolute secrecy rate is a highly pessimistic measure for secrecy. Next, we introduce a general statistical secrecy measure which includes the absolute secrecy rate as a special case. In particular, we quantify the \textit{probability} that information is leaked to Eve when the transmitter transmits with a fixed rate $R_s$ [bits/s/Hz]. Therefore,  the absolute secrecy rate corresponds to the special case when the probability of information leakage  is zero. 

For a given rate $R_s$, information leakage probability, denoted by $P_{\rm leak}$, can be formally defined as
\begin{align}\label{Eq:Pleak}
	P_{\rm leak} = \Pr\left\{C_{s}(\bH_e) \leq R_s \right\}, 
\end{align}
where $C_{s}(\bH_e)$ is the instantaneous achievable secrecy rate which is given by
\begin{align}\label{Eq:Rs_ins}
	C_{s}(\bH_e) \!= \!\!\underset{\bp_e\in\Pset_e}{\min} &\sum_{l=1}^{L} \! T_l \big[C_{b,l}-C_{e,l}(\bH_e)\big]^+ 
\end{align}
    where $\bH_e\in\Hset_{e}(\bp_e)$. Note that the key difference between the achievable absolute  secrecy rate $\bar{C}_{s}^{\rm abs}$ in \eqref {Eq:Rs_abs} and the instantaneous achievable secrecy rate $C_{s}(\bH_e)$ in \eqref{Eq:Rs_ins} is that in \eqref {Eq:Rs_abs}, the worst-case realization is assumed for $\bH_e$, whereas in \eqref{Eq:Rs_ins}, the actual (unknown) realization is assumed.   


\subsection{Wireless channel knowledge}\label{Sec:TimeAllocation}
To ensure reliable decoding at Bob, BeamSec requires that the quality of the legitimate link, i.e., $\gamma_{b,l},\,\,\forall l$, to be known at Alice for choosing $R_s< \sum_{l=1}^{L} \! T_l \log_2(1+\gamma_{b,l})\triangleq C_b$. This information is obtained through beam training. BeamSec does not require any instantaneous or statistical knowledge of the eavesdropper's channel or its location. In fact, BeamSec works for any choices of the time assignments $T_l,\,\,\forall l,$ and the transmission rate $R_s< C_b$. Nonetheless, since the eavesdropper(s) and the legitimate \gls{rx} access the same wireless channel, the empirical channel measurements collected by the legitimate \gls{rx} over time provide useful knowledge that can be exploited for optimizing the values of $T_l,\,\,\forall l,$ and $R_s$. 
To show this, we study two options: $(i)$  No knowledge about the eavesdroppers or the wireless channel is available.  $(ii)$ Only partial knowledge of the wireless channel is available, which is collected through sparse empirical measurements by legitimate \gls{rx} in the past. 

\subsubsection{No knowledge of wireless channel}
Here, we allocate the same time duration for each beam-pair, i.e., $T_l=\frac{1}{L},\,\,\forall l.$
This strategy does not require knowledge regarding the wireless channel and the change in the codeword length. Since the value of $\bar{C}_{s}^{\rm abs}$ is unknown, one cannot ensure zero information leakage (i.e., valid when $R_s<\bar{C}_{s}^{\rm abs}$). Therefore, information leakage probability is the proper performance measure.

\subsubsection{Partial channel knowledge}
\label{ss:Par_knowledge}In this case, we assume that the eavesdropper experiences the same channel statistics at location $\bp_e$ as those measured empirically by the legitimate \gls{rx} at location $\bp_e$. Therefore, despite the unavailability of the current location of the eavesdropper, an estimate of $\bar{C}_{s}^{\rm abs}$ can be computed from \eqref{Eq:Rs_abs} based on the  past empirical channel measurements (which can be available only for a subset of locations $\hat{\Pset}_e\subset\Pset_e$). Alternatively, the partial channel knowledge can be exploited to compute an estimate of $P_{\rm leak}$ from \eqref{Eq:Pleak}.Furthermore, with the above partial knowledge of the wireless channel,  one can optimize $T_l,\forall l$, to maximize the secrecy rate. In particular, the optimization problem for maximizing the achievable absolute secrecy rate $\bar{C}_{s}^{\rm abs}$ in terms of time variable $T_l$ can be formulated as follows 
\begin{align} \label{Eq. optimum time allocation}
	&\underset{T_l,\forall l,\,\,t}{\rm maximize} \quad \underset{\bp_e\in\hat{\Pset}_e}{\min} \sum_{l=1}^{L} T_l C_{s,l}(\bp_e) \nonumber\\
	&{\rm subject\,\,to}\quad 0\leq T_l\leq 1,\,\,\forall l, \,\,\text{and}\,\,\sum_{l=1}^{L}T_l=1.
\end{align}
Defining auxiliary optimization variable $t$ for the epigraph of the cost function, we can transform the above problem into 
\begin{align}\label{Eq:TlProb}
	&\underset{T_l,\forall l}{\rm maximize} \quad t  \nonumber\\
	&{\rm subject\,\,to}\quad \sum_{l=1}^{L} T_l C_{s,l}(\bp_e)\geq t,\quad \forall \bp_e\in\hat{\Pset}_e
	\nonumber\\
	&\qquad\qquad\quad 0\leq T_l\leq 1,\,\,\forall l, \,\,\text{and}\,\,\sum_{l=1}^{L}T_l=1.
\end{align}
The above optimization problem becomes linear programming with $|\hat{\Pset}_e|+L_b+1$ linear constraints, which can be solved using standard numerical solvers for convex optimization, e.g., CVX. 

Alternatively, for a given fixed rate $R_s$, one can, in principle optimize $T_l,\forall l$, in order to minimize $P_{\rm leak}$. However, the formulation of the corresponding optimization problem requires analytical characterization of $P_{\rm leak}$ in \eqref{Eq:Pleak}, which is beyond the scope of this paper but constitutes an interesting direction for future research. Nevertheless, in Section \ref{Sec:Experiment}, we show that the values of $T_l,\,\,\forall l$, obtained from \eqref{Eq:TlProb} for maximizing $\bar{C}_{s}^{\rm abs}$ can still significantly improve $P_{\rm leak}$ compared to baseline uniform time allocation, i.e., $T_l=\frac{1}{L},\,\,\forall l$.

  \begin{figure}[t]
\centering
\begin{minipage}{0.47\linewidth}
\includegraphics[width=1\linewidth,keepaspectratio]{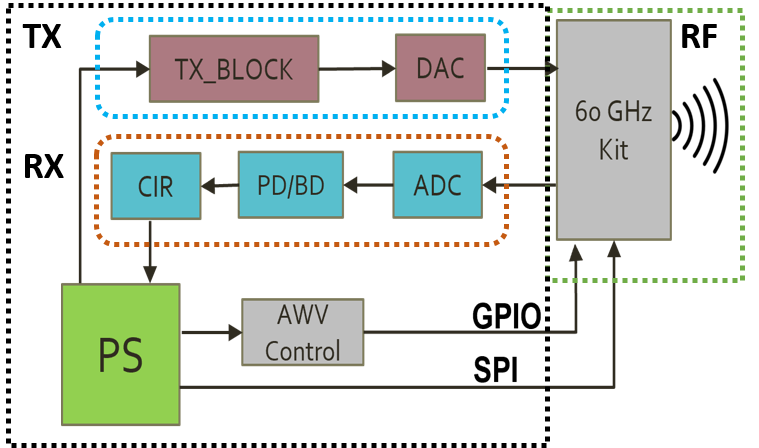}
\end{minipage}
\begin{minipage}{0.47\linewidth}
\includegraphics[width=1\linewidth,keepaspectratio]{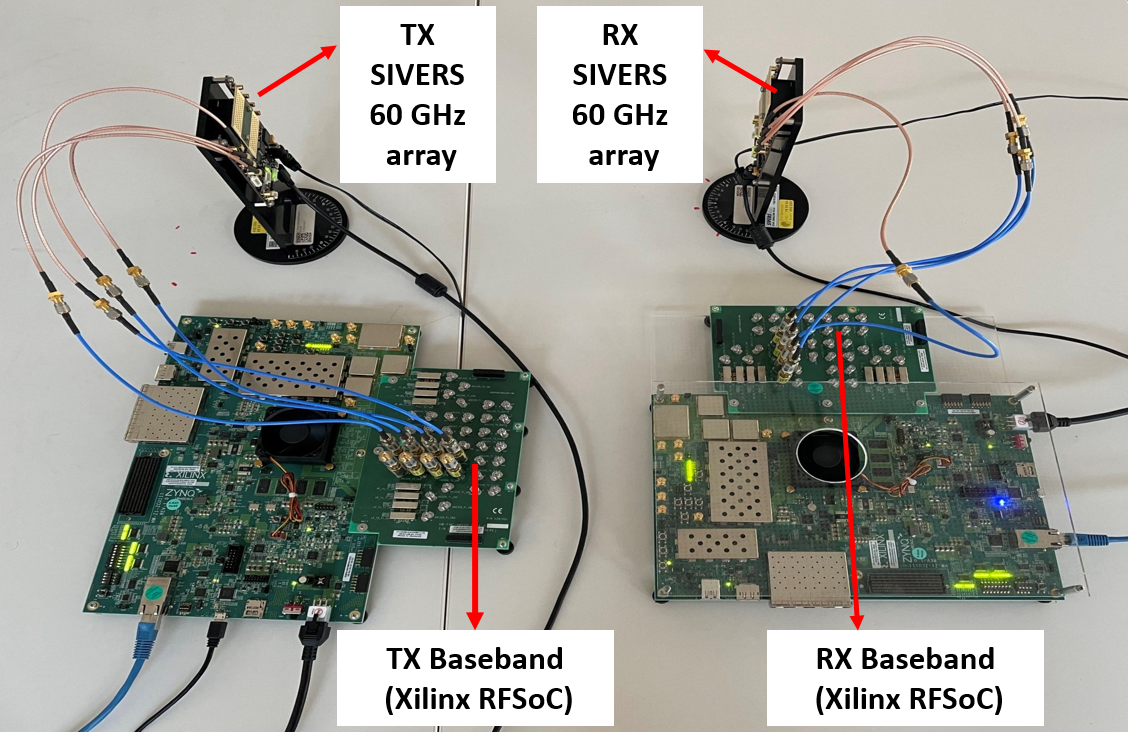}
\end{minipage}
\caption{An overview of our 60GHz testbed including Baseband (Xilinx RFSoC) and RF transceiver (60GHz phased arrays).}
\label{fig_TxRxLoop}

\end{figure}

\section{Experimental Setup}\label{Sec:Experiment}
\begin{figure}[!t]
    \centering
    \includegraphics[scale=0.23]{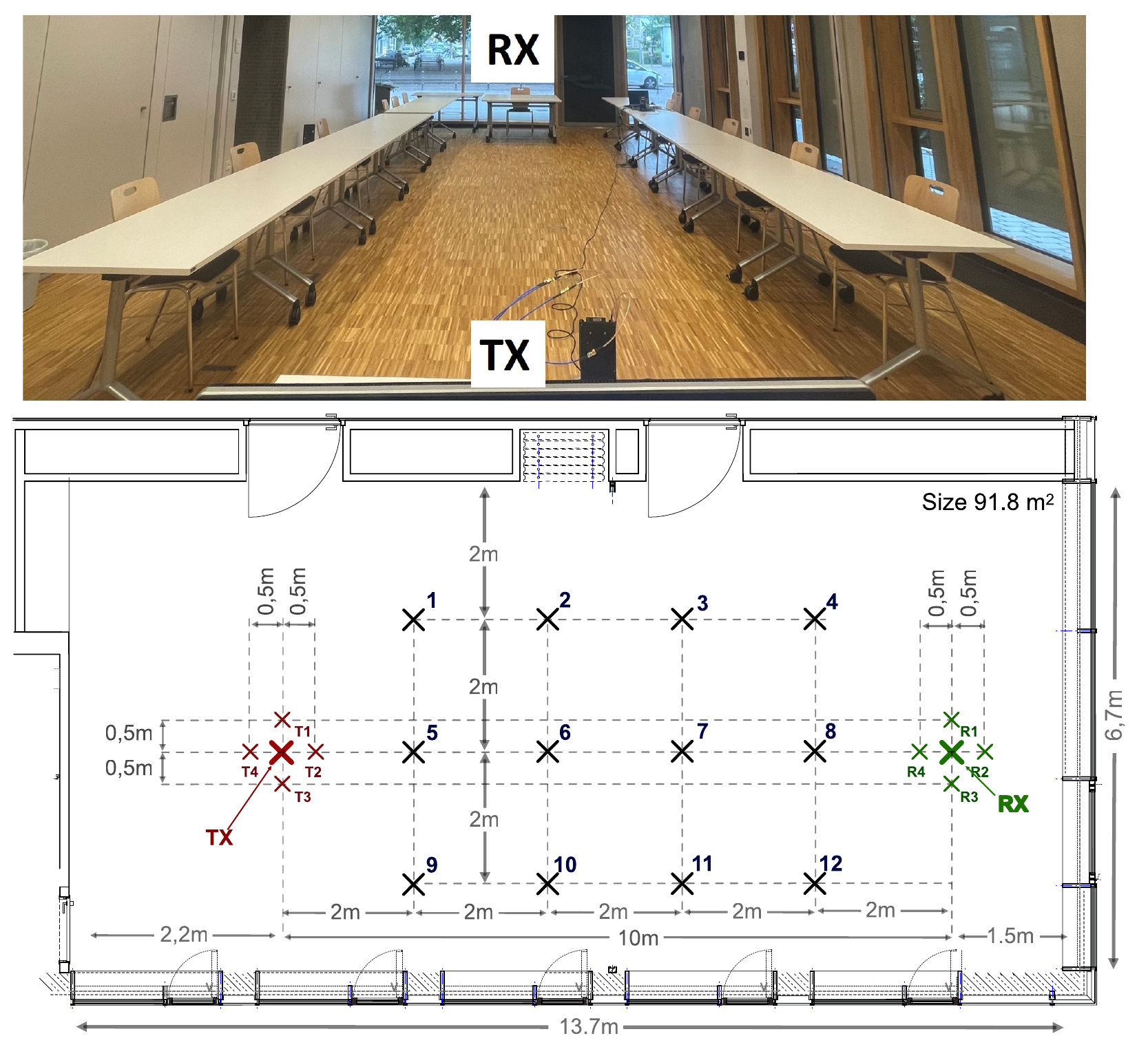}%
    \caption{Overview and layout of the seminar room and the exact location of measurement points.}
    \label{fig: Distancemap_Eve}
\end{figure}

 We deploy \acrshort{BSec} on an \gls{sdr}-based testbed using Xilinx RFSoC ZCU111 and SiversIMA 60 GHz RF front-end with 2 GHz bandwidth (\figref{fig_TxRxLoop}). The RF front-end includes a 16-element phased-array antenna for analog beamforming. For the software, we use the open-source implementation of 802.11ad/ay available at~\cite{23}. The legitimate communication parties (Alice and Bob) synchronize according to the 802.11ad procedure. The beam switching occurs at intervals agreed upon after the optimization phase. To model a strong adversary, Eve is synchronized to Alice---a practical Eve would either have to feature advanced synchronization capabilities or suffer a higher packet loss probability. In a 91.8~$m^2$ seminar room, our experimental configuration includes typical indoor elements, like metal panels, concrete walls, and glass windows, along with a 75~$cm$ high antenna. Experiments were conducted during office hours in an unoccupied environment. Our phased array offers an angular span of $\pm 45^\circ$, extendable to $\pm 135^\circ$ through antenna rotation. \figref{fig: Distancemap_Eve} displays the room layout, seminar space, and varied Eve locations. We systematically evaluated \acrshort{BSec} against diverse Eve attack models: $(i)$ Quasi-omni, $(ii)$$ Directional, and $(iii) Colluding. Transmission duration ($T$) is set at 100 packets, optimally/uniformly partitioned among paths, with 100-packet repetitions per path, resulting in $L*100$ packets per location.
\begin{figure}[t!]
    \centering
   \begin{subfigure}[h]{.47\linewidth}
         \centering
         \includegraphics[width=\linewidth]{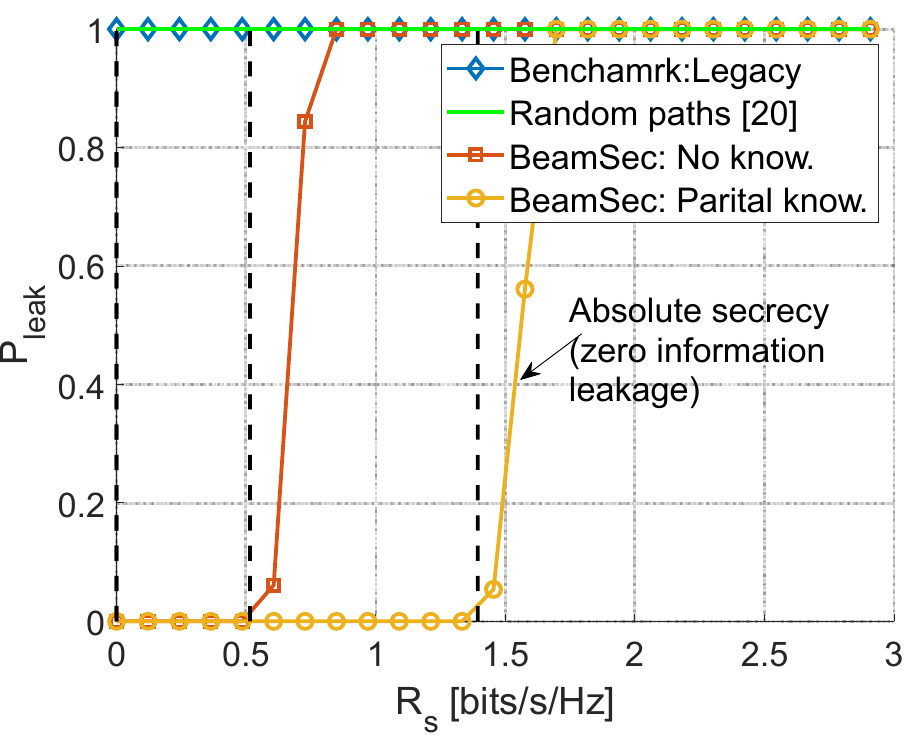}  \caption{\acrshort{att1} }
     \end{subfigure}
     \hfill
     \begin{subfigure}[h]{.47\linewidth}
         \centering
         \includegraphics[width=\linewidth]{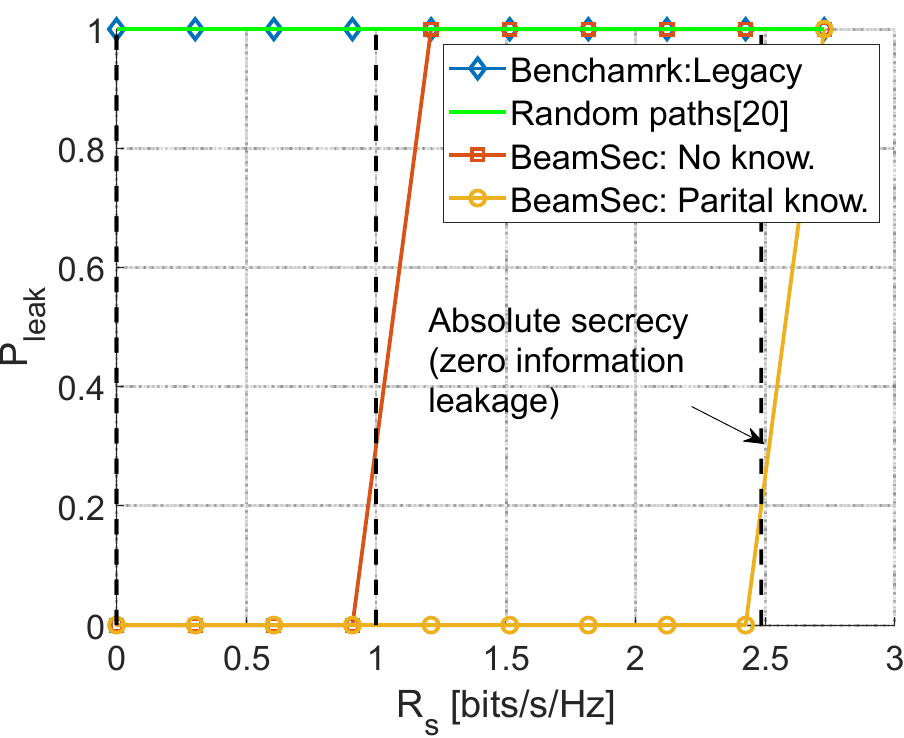} \caption{\acrshort{att2}}
         
     \end{subfigure}
  \caption{Information leakage probability versus the transmission rate for the single attacker scenario. }
        \label{fig:Pleak}
\end{figure}

\section{Experimental Evaluation}
In this section, we perform extensive experiments to evaluate \acrshort{BSec}’s performance in an indoor scenario, covering single and colluding eavesdropper scenarios.

\subsection{Single eavesdropper}
\label{single eavesdropping attack}
\subsubsection{Information leakage probability}
In \figref{fig:Pleak}, we demonstrate the probability of leakage, $P_{\rm leak}$, versus the transmission rate, $R_s$,  under \acrshort{BSec} with no/partial knowledge,
legacy approach (i.e., choosing the best beam) and random path hopping~\cite{16}. The absolute secrecy rate, $\bar{C}_s^{\rm abs}$, can be inferred from the leakage probability by identifying the maximum rate $R_s$ for which $P_{\rm leak}$ is zero. First, we observe from \figref{fig:Pleak} that the legacy approach leads to nearly complete leakage even for very low data rates. This emphasizes the fact that the \gls{los} link is highly exposed to the attackers, in particular directional attackers, which benefit from higher directional antenna gain. The high leakage stems from the exposure of the main-lobe and unwanted side-lobe and scatterers along their path.  Secondly, random paths also have $P_{\rm leak}=1$ as the paths have limited diversity and might feature minimum angular separation in \gls{aod}, which leads to capture all parts of data transmission by Eve. In contrast, \acrshort{BSec} provides a non-zero secure rate (with zero information leakage) for both attacker models. In particular, for a \acrshort{att1}, \acrshort{BSec} with zero knowledge provides an absolute secure region with zero leakage if $R_s \leq 0.55$ [bits/s/Hz].  \acrshort{BSec} with partial knowledge shifts the absolute secure region to $1.39$ [bits/s/Hz] (i.e., improved by $\sim\!40\%$). For a \acrshort{att2}, \acrshort{BSec} with zero knowledge provides an absolute secure region with zero leakage if $R_s \leq 1$ [bits/s/Hz]. For \acrshort{BSec} with partial knowledge, which further optimizes the paths based on their worst channel realizations, a considerable increase in the absolute secure region can be observed with a higher secrecy rate, i.e., $R_s\leq 2.52$ [bits/s/Hz]  (i.e., $\sim\!151\%$ improvement compared to uniform time allocation). 
 


\begin{figure}[t!]
     \centering
     \hspace{-10mm}
     \begin{subfigure}[h] {.22\linewidth}
         \centering
         \includegraphics[trim={0 0 5cm 0},clip,width=1.1\linewidth]{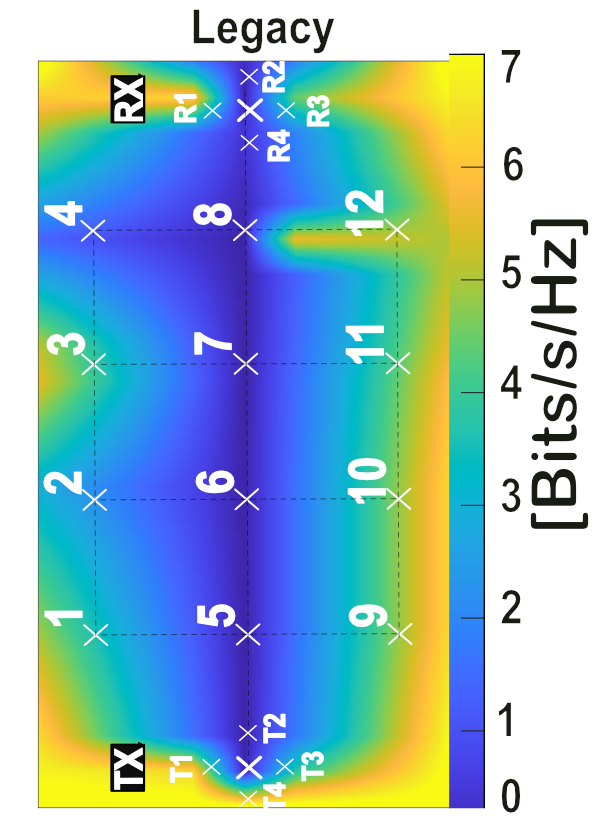}
         \caption{}
            
         \label{SR_OmniCPA}
     \end{subfigure}
     \begin{subfigure}[h]{.22\linewidth}
         \centering
         \includegraphics[trim={0 0 5cm 0},clip,width=1.1\linewidth]{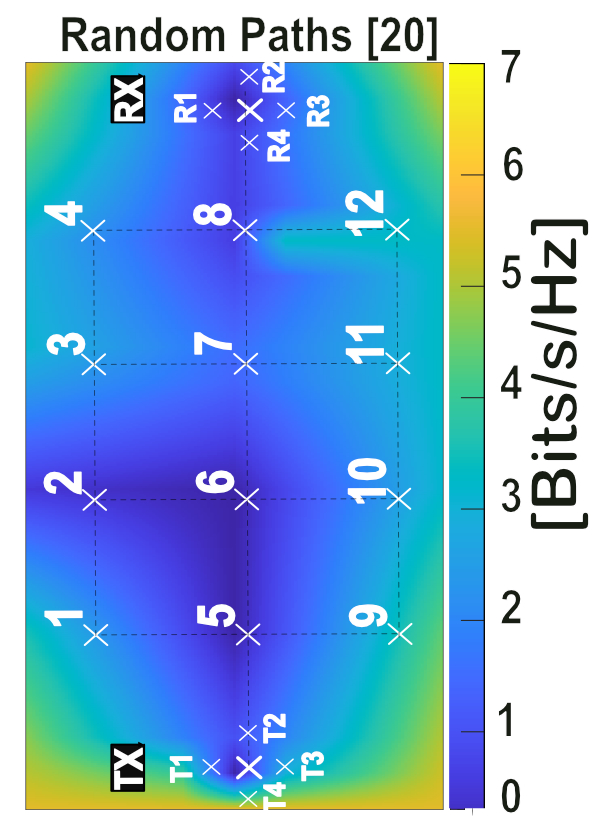}
         \caption{}
         
         \label{SR_OmniRnd}
         
     \end{subfigure}
     \begin{subfigure}[h]{.22\linewidth}
         \centering
         \includegraphics[trim={0 0 5cm 0},clip,width=1.1\linewidth]{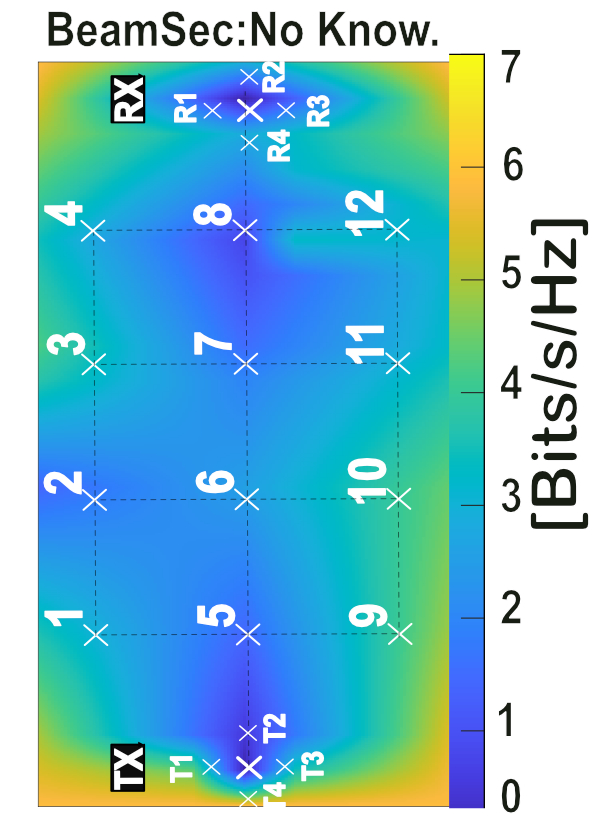}
            \caption{}
         
         \label{SR_OmniUnif}
     \end{subfigure}
     \begin{subfigure}[h]{.22\linewidth}
         \centering
         \includegraphics[trim={0 0 0 0},clip,width=1.43\linewidth]{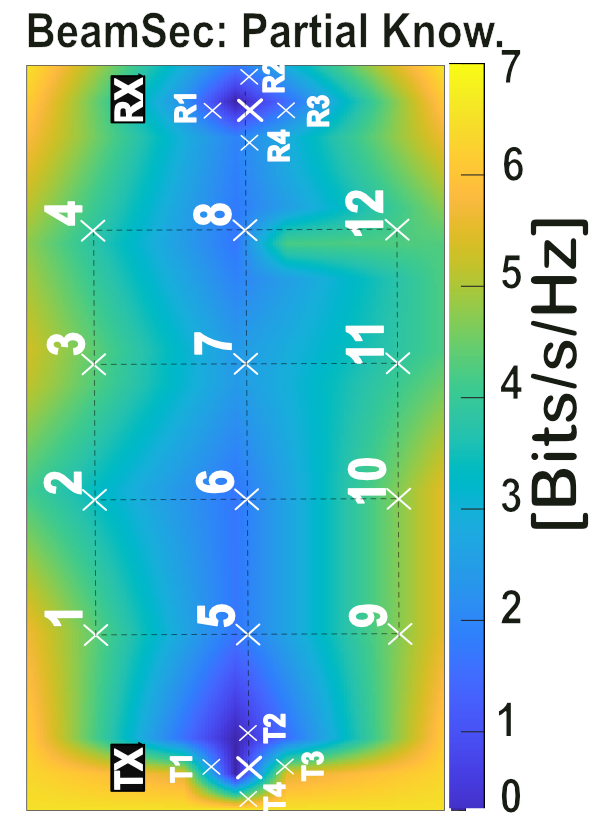}
         \caption{}
         \label{SR_OmniOpt}
     \end{subfigure}

         \caption{Heatmap of the absolute secrecy rate $\bar{C}_s^{\rm abs}$ for \acrshort{att1}. By transmitting along multiple beams, \acrshort{BSec} reduces the signal footprint and hence improves the secrecy rate. On average, over all locations,  \acrshort{BSec} with partial knowledge performs $~79.8$\% better than random~\cite{16}.}
        \label{fig:SR1}
\end{figure}  
\subsubsection{Absolute secrecy rate}

In this section, we discuss the experimental results for absolute secrecy $\bar{C}_s^{\rm abs}$  for different locations of Eve (unknown to Alice and Bob\footnote{For better visualization, we have interpolated the data points to observe the performances in a heatmap.}). 

\textbf{The \acrshort{att1}:}
\figref{fig:SR1} shows the heatmap for the secrecy rate, which is obtained by the interpolation of  the measurement points provided in \figref{fig: Distancemap_Eve}  for the benchmark legacy (LoS) and random paths selection, \acrshort{BSec} with no/partial knowledge of the wireless channel. Due to less directionality gain of quasi-Omni,  we expect to capture \gls{los} only at \textit{Locs 5,6,7,8} through its main-lobe, due to which the secrecy rate in \gls{los} is zero, as can be seen in \figref{SR_OmniCPA}. As other locations do not intercept with high quality, the  secrecy rate is non-zero at these locations. However, since Eve's location is unknown, the worst-case location is the bottleneck for determining secure communication, and therefore, the overall absolute secrecy rate of the considered benchmark is zero. Random paths selection chooses a subset of paths from a set of paths reaching RX randomly, which leads to the selection of closely correlated paths; hence the advantage of using multiple paths is diminished. Data transmission over correlated paths results in capturing the whole transmission, e.g., \textit{Locs 2,5,6} in \figref{SR_OmniRnd} where the absolute secrecy rate is zero.
In contrast, \acrshort{BSec} provides a non-zero secrecy rate with absolute secrecy by exploiting the path diversity. With no knowledge of Eve, the worst-case Eve's location is \textit{Loc 8}, which limits the absolute secrecy rate to $0.76$~[bit/s/Hz] (\figref{SR_OmniUnif}). With partial channel knowledge, the worst-case Eve's location becomes \textit{Loc 8}, and the absolute secrecy rate increases to $1.70$~[bit/s/Hz] (\figref{SR_OmniOpt}). 

\begin{figure}[t!]
 \centering
 \hspace{-9mm}
     \begin{subfigure}[h]{.22\linewidth}
         \centering
         \includegraphics[trim={0 0 5cm 0},clip,width=1.1\linewidth]{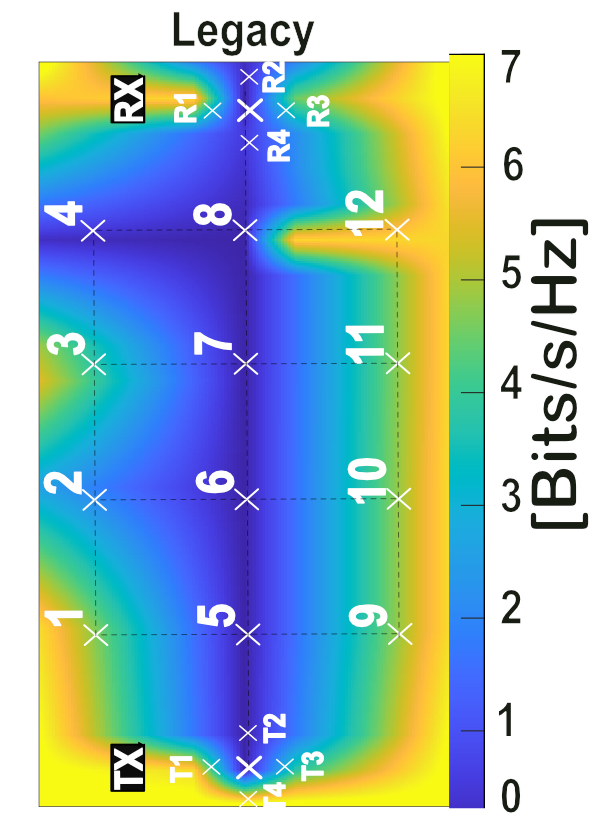}         
         \caption{}
         \label{SR_DirecCPA}
     \end{subfigure}
     \begin{subfigure}[h]{.22\linewidth}
         \centering
         \includegraphics[trim={0 0 5cm 0},clip,width=1.1\linewidth]{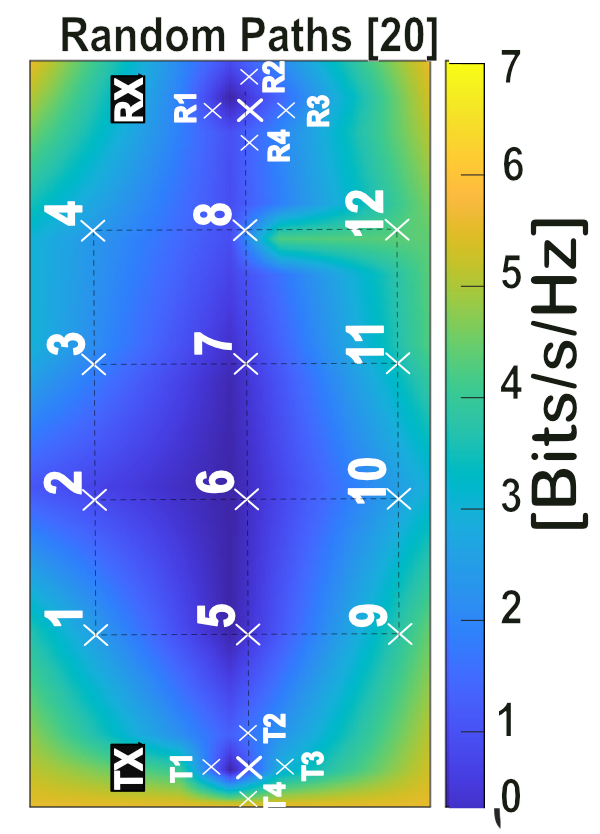}
             \caption{}
         
         \label{SR_RndDR}
         
     \end{subfigure}
     \begin{subfigure}[h]{.22\linewidth}
         \centering
  \includegraphics[trim={0 0 5cm 0},clip,width=1.1\linewidth]{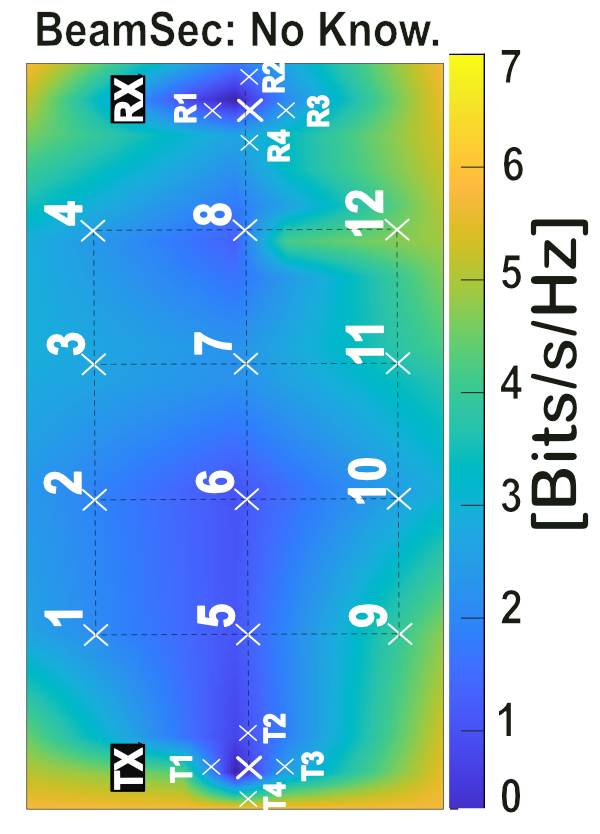}
             \caption{}   
         \label{SR_DirecUnif}
     \end{subfigure}
     \begin{subfigure}[h]{.22\linewidth}
         \centering
         \includegraphics[width=1.43\linewidth]{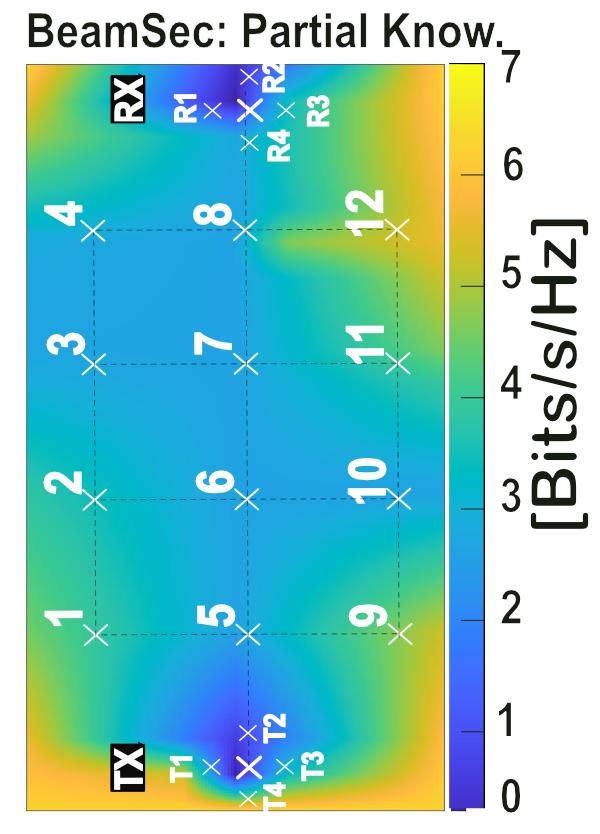}
             \caption{}
         
         \label{SR_DirecOpt}
         
     \end{subfigure}
         \caption{Heatmap of the absolute secrecy rate $\bar{C}_s^{\rm abs}$ for \acrshort{att2}. On average, over all locations,  \acrshort{BSec} with partial channel knowledge yields $~59.7$\% improvement in secrecy rate  compared to random path selection.}  
        \label{fig:SR2}
\end{figure}

\textbf{The \acrshort{att2}:}
A \acrshort{att2} considers eavesdroppers with strong capabilities such as perfect synchronization (zero-time) and reception on an optimal beam (high quality). \figref{SR_DirecCPA} shows that the secrecy rate is comparatively reduced as compared to \acrshort{att1} (\figref{SR_OmniCPA}) for all schemes.  Due to high receiver sensitivity and directional gain,  the secrecy rate for the legacy is reduced at nLoS locations (i.e., zero at \textit{Loc 4}). The reason is that the nLoS locations can still receive the signal emitted through side-lobes or reflections. Random paths in \figref{SR_RndDR}) also show zero secrecy rate for \textit{Loc 2,5,6,7,8} \acrshort{BSec} with no knowledge of Eve provides a non-zero secrecy rate at all locations; however, the secrecy rate significantly drops at \textit{Loc 5}, which limits the overall secrecy rate to $1.064$ [bits/s/Hz] (\figref{SR_DirecUnif}). This location is closer to TX and makes \gls{los} with both side-lobes, and main-lobes vulnerable to eavesdropping. \acrshort{BSec} with partial knowledge of Eve provides absolute secrecy rate of $2.64$ [bits/s/Hz]. 

\subsubsection{Impact of channel knowledge}
\begin{figure}[t]
 \centering    
       \includegraphics[scale=0.55,keepaspectratio]{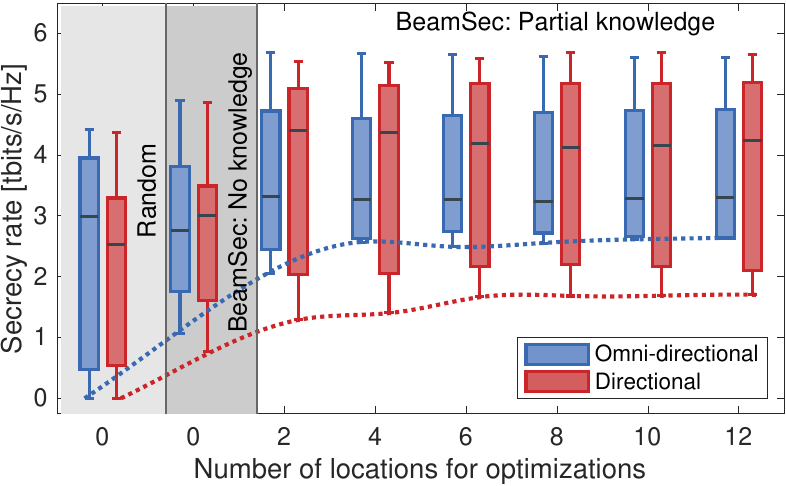}
    \caption{Secrecy rate versus the number of measurement points for optimization.}
    \label{fig: SecrecyRate_Opt_Locations}
\end{figure}

In \figref{fig: SecrecyRate_Opt_Locations}, the secrecy rate is illustrated based on the number of measurement points used for the optimization problem in \eqref{Eq. optimum time allocation}. As mentioned, the absolute secrecy rate is the minimum achievable secrecy rate obtained in worst-case channel realization. It is observed that increasing the number of measurement points (channel knowledge) leads to an upward trend in the absolute secrecy rate, which is shown in red and blue dashed lines for the omni-directional and directional eavesdroppers, respectively.
     

        

\begin{figure}[b]
 \centering
     \begin{subfigure}[h]{0.48\linewidth}
         \centering
         \includegraphics[width=\linewidth]{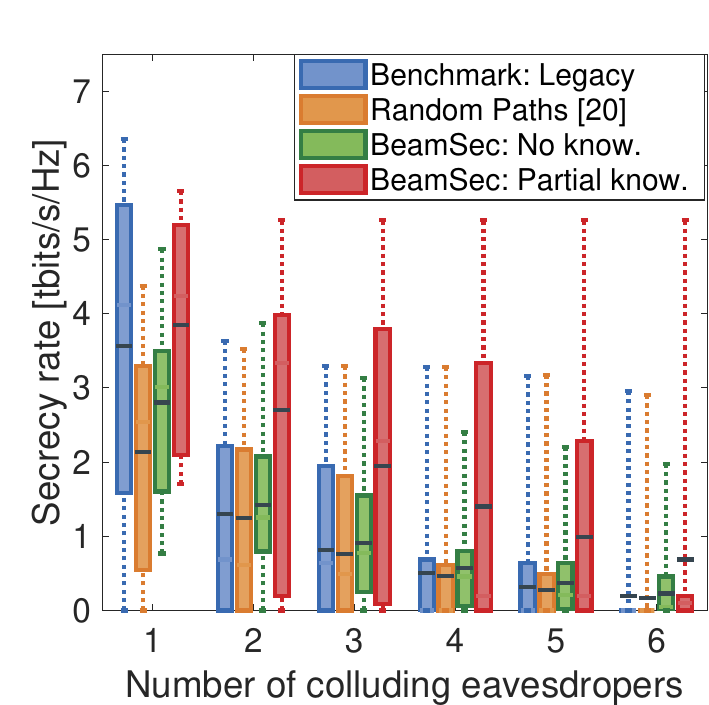}
            
         \caption{\acrshort{att1}}
         \label{SR_ColludOmni}
     \end{subfigure}
     \begin{subfigure}[h]{0.48\linewidth}
         \centering
         \includegraphics[width=\linewidth]{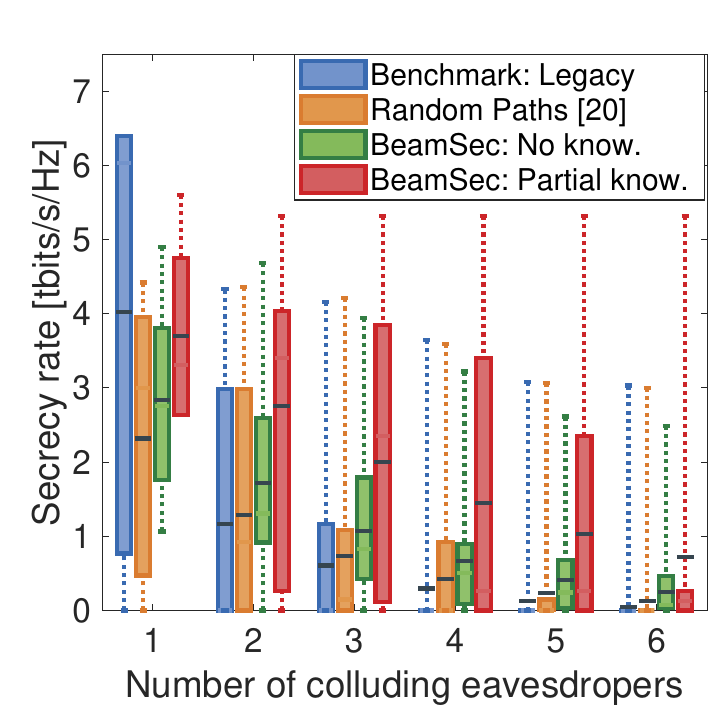}
             \caption{\acrshort{att2}}
         
         \label{SR_ColludNom}
     \end{subfigure}

         \caption{Secrecy rate for colluding eavesdroppers.}
        
        \label{fig:SRCollud}
\end{figure}

 \subsection{Colluding eavesdroppers}
 One of the unique features of \acrshort{BSec} is resilience toward coordinated and colluding eavesdroppers. To better represent this, we have computed the average result of worst channel realizations w.r.t. secrecy rate of all the possible combinations of colluding eavesdroppers in Figs~\ref{SR_ColludOmni}~and~\ref{SR_ColludNom}, illustrating the impact of the number of colluding eavesdroppers on the secrecy rate. As this number rises, the absolute and average secrecy rates decline. To present a complete depiction, \figref{fig:SRCollud} includes the case of a single eavesdropper, which represents a non-colluding case. For \acrshort{att1}, we observe that \acrshort{BSec}-partial knowledge of wireless channel provides the highest secrecy rate in all cases. While the secrecy rate of the legacy approach drops to nearly $0.1$ with six colluding eavesdroppers, \acrshort{BSec}-partial knowledge of the channel maintains $0.7$ [bits/s/Hz]. The trend also shows that other schemes are  more vulnerable to the multi-attacker scenarios, as for \acrshort{att1}, the secrecy rate drops by $64.2$\% for legacy and $42$\% for random paths selection with only two attackers, whereas \acrshort{BSec}-partial knowledge of channel only experiences a $29$\% drop. While Figs \ref{fig:Pleak} and \ref {fig:SR1} revealed that \acrshort{BSec}-no channel knowledge outperforms the legacy scheme in terms of the worst-case secrecy rate over all locations, \figref{SR_ColludOmni} indicates \acrshort{BSec}-no channel knowledge's lower average secrecy rate than legacy with one eavesdropper. To ensure a comprehensive comparison, it's necessary to examine not just worst-case and average performance but the entire achievable secrecy rate statistics for all colluding combinations.

\section{Discussions}


\subsection{Reducing the impact of side-lobes}
During measurements, it was observed that side-lobes impact can be suppressed only to a certain level. 
The performance of \acrshort{BSec} can be further improved by devising a codebook with a high side-lobe suppression level. While this is out of the scope of this work, PLS-focused beamforming and antenna design are interesting future research directions.

\subsection{Impact of environments on the number of distinct paths}

\begin{figure}[t]
 \centering
 
     \begin{subfigure}[h]{.89\linewidth}
         \includegraphics[width=0.89\linewidth]{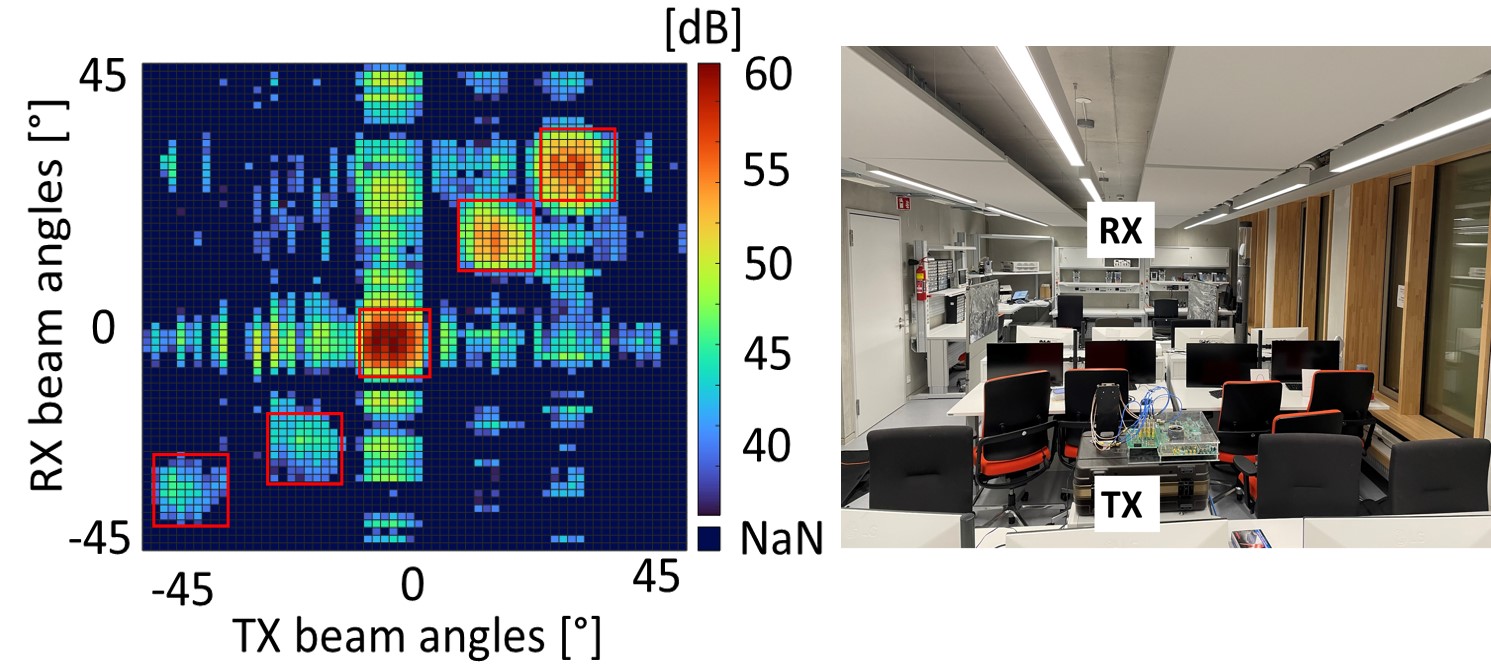}
          \vspace{-3mm}
         \caption{Hardware lab}
         \label{subfig:a}
     \end{subfigure}\\
 
     \begin{subfigure}[h]{.89\linewidth}
         \includegraphics[width=0.89\linewidth]{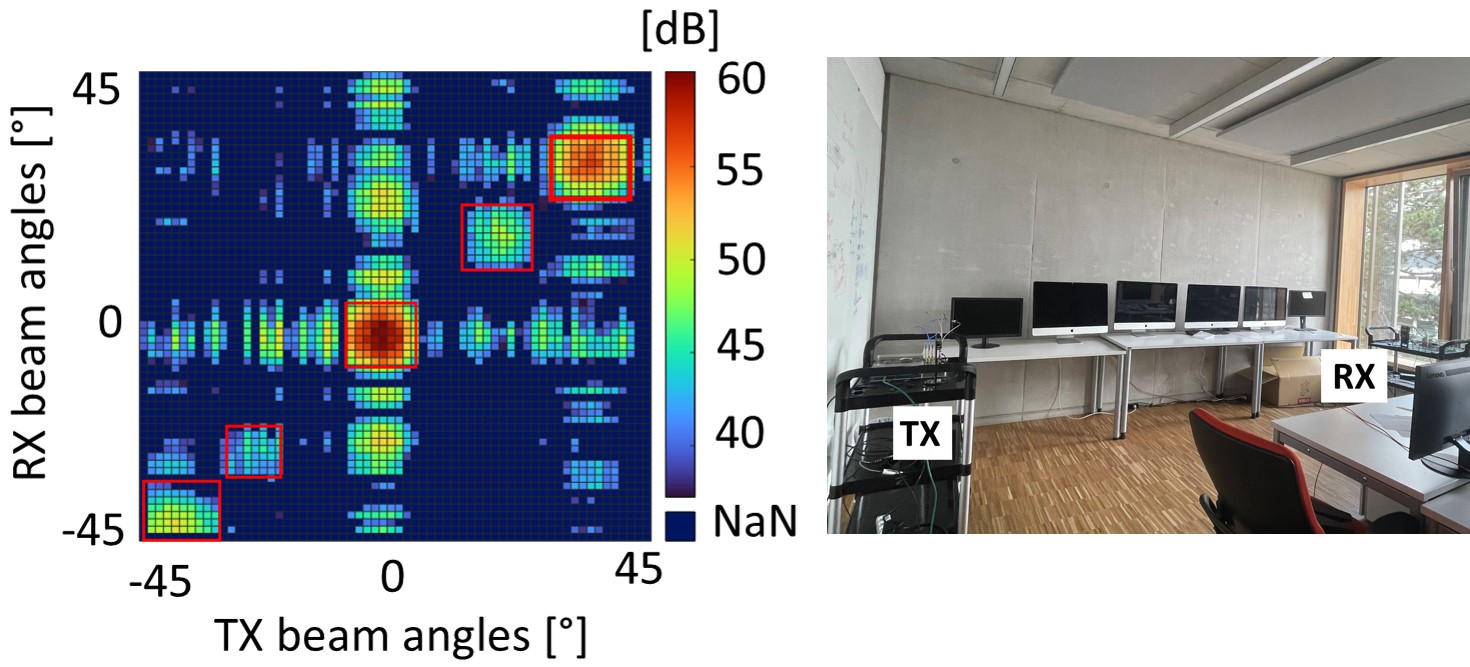}
          \vspace{-3mm}
         \caption{Computer lab}
         \label{subfig:b}
     \end{subfigure}
         \caption{Environment characterization.}       
        \label{fig: Env_Ch}
\end{figure}

\acrshort{BSec}'s effectiveness depends on multiple path diversity, influenced by the surroundings (e.g., furniture, building material). We tested this in two additional areas: $(i)$ a cluttered hardware lab (61 $m^2$) and $(ii)$ an open office-like computer lab (24 $m^2$). \figref{fig: Env_Ch}  shows room layouts and their channel profiles. Both cases showed at least two distinct nLoS paths and one \gls{los} path, potentially increased with both azimuth-elevation sweeping. Elevation scans use ceiling-floor reflections. Alternatively, antennas with tunable polarization can be used when the environment is not reflective (e.g., a wooden cabin).

\subsection{Controlling the Environment}
While channel propagation remains uncontrollable, deploying reconfigurable intelligent surfaces (RISs) indoors has been explored to enhance wireless communication. RISs can likely boost \acrshort{BSec} performance by generating extra paths and refining signal focus, potentially minimizing the footprint. Exploring effective coordination among TX, RX, and RIS is a promising research direction.

\section{Conclusion} 
 We propose a practical main-lobe security scheme for mmWave systems. Specifically, we select a set of diverse and distinct beam-pairs by analyzing the environmental conditions and the signal characterization. Furthermore, we devise two methods for time allocation among different beam-pairs. The result of the experimental evaluation indicates that \acrshort{BSec} can significantly improve the secrecy rate of mmWave systems in the presence of both single and colluding eavesdroppers.


\section{Acknowledgements}
Funded by the European Commission’s Horizon 2020 through Marie Skłodowska-Curie Action MINTS (GA no. 861222), Waqar Ahmed’s work is supported by DFG within SFB 1053 MAKI, while Khaloopour and Jamali receive partial support from LOEWE initiative via emergenCITY center.


{\footnotesize \bibliographystyle{IEEEtran}
\bibliography{CNS.bib}}

\end{document}